\documentclass[12pt]{article}
\usepackage{graphicx}
\newcommand{\dbox}{\,\raise2pt\hbox{\fbox{\rule{2.5pt}{0pt}\rule{0pt}{2.5pt}}}\,}
\newcommand{\qed}{\,\raise0pt\hbox{\mbox{\rule{6.5pt}{6.5pt}}}}
\newcommand{\bra}[1]{\mbox{$\langle #1 |$}}
\newcommand{\ket}[1]{\mbox{$| #1 \rangle$}}

\newtheorem{lemma}{Lemma}
\newtheorem{theorem}{Theorem}

\begin{document}
\setlength{\baselineskip}{7mm}

\begin{titlepage}
 \begin{normalsize}
  \begin{flushright}
        UT-Komaba/08-12\\
        July 2008
  \end{flushright}
 \end{normalsize}
 \begin{LARGE}
   \vspace{1cm}
   \begin{center}
General Linear Gauges and Amplitudes\\ in Open String Field Theory \\
   \end{center}
 \end{LARGE}
  \vspace{5mm}
 \begin{center}
    Masako {\sc Asano} 
            \hspace{3mm}and\hspace{3mm}
    Mitsuhiro {\sc Kato}$^{\dagger}$ 
\\
      \vspace{4mm}
        {\sl Faculty of Liberal Arts and Sciences}\\
        {\sl Osaka Prefecture University}\\
        {\sl Sakai, Osaka 599-8531, Japan}\\
      \vspace{4mm}
        ${}^{\dagger}${\sl Institute of Physics} \\
        {\sl University of Tokyo, Komaba}\\
        {\sl Meguro-ku, Tokyo 153-8902, Japan}\\
      \vspace{1cm}

  ABSTRACT\par
 \end{center}
 \begin{quote}
  \begin{normalsize}
A general class of gauges for open string field theory, for which the gauge fixing condition is given by linear equations with respect to string field, is investigated in detail. This class of gauges includes almost all known ones like Siegel gauge and its various extensions such as $a$-gauges proposed by the present authors as well as Schnabl gauge and linear $b$-gauges. A general form of propagators is determined and their common features are analyzed. A consistent procedure for calculating the amplitudes is given. Gauge independence of the on-shell physical amplitudes is explicitly shown.
\end{normalsize}
 \end{quote}

\end{titlepage}
\vfil\eject

\section{Introduction}

Recent rapid developments on the analytic solutions in open string field theory~\cite{Schnabl:2005gv}-\cite{Kawano:2008ry} are triggered by Schnabl's analysis~\cite{Schnabl:2005gv} of tachyon vacuum where he utilized a new gauge which greatly simplifies computations involving Witten's star-product~\cite{Witten:1985cc} of string fields. On the other hand, Siegel gauge, which has been known since early days of string field theory, is in conformity with the worldsheet picture, and it naturally fits into perturbative calculations. This suggests that the choice of gauge may become crucial in the analysis of concrete individual problems.

The present authors proposed in ref.\cite{Asano:2006hk} one-parameter family of gauges, which we refer $a$-gauges throughout the present paper since we use $a$ as a gauge parameter. A special point $a=0$ is equivalent to Siegel gauge.
For massless vector mode, this $a$-gauges naturally reduce to the covariant gauges with a gauge parameter in ordinary gauge theories including Landau gauge and Feynman (Siegel) gauge. As in the ordinary gauge theory, a suitable choice of the gauge parameter may give a useful simplification in distinct applications. Indeed, $a$-gauges gave alternatives to Siegel gauge in the level truncation analysis~\cite{Asano:2006hm} and was utilized to efficiently distinguish physical branch from gauge artifact in tachyon potential.

Another generalization of known gauges is linear $b$-gauges studied in ref.\cite{Kiermaier:2007jg}, which is useful for regularizing a singular behavior of the propagator in Schnabl gauge for the calculation of perturbative amplitudes~\cite{Kiermaier:2008jy, Rastelli:2007gg}.

In the present paper we investigate properties of the perturbative amplitudes in 
more general class of gauges which we call general linear gauges, where we mean by {\it linear} that the gauge fixing condition is linear in string field, but the operator acting on the field is not restricted to be linear. Schematically gauge fixing condition for a string field $\Phi$ is given by
\begin{equation}
{\cal A} \Phi = 0
\end{equation}
with some operator ${\cal A}$. This class of gauges is so general as to contain almost all known gauges such as Siegel gauge (${\cal A}$ is just $b_0$) and its extensions like $a$-gauges (${\cal A}$ will be given in the succeeding section), Schnabl gauge and linear $b$-gauges (${\cal A}$ is a linear combination of $b_n$'s). 
String field theories in this class of gauges have many common features, e.g. the form of the propagators and the identities they satisfy, which will play important role in the proof of gauge independence of on-shell physical amplitudes.

In the next section we will construct the gauge fixed action in general linear gauges and will show a general prescription for deriving the propagators for arbitrary ghost number sector. We also collect the general properties of the propagators.
As an example, we explicitly calculate the propagators for $a$-gauges and linear $b$-gauges. 
In section three we will give a consistent procedure for defining the perturbative amplitudes. Then we show the gauge independence of the on-shell physical amplitudes in section four. The final section will be devoted to summary and discussions.

\section{General linear gauges and the gauge fixed action}

In this section, we analyze the general perturbation for 
the cubic open string field theory~\cite{Witten:1985cc} 
whose action is given by 
\begin{equation}
S = -{1 \over 2} \langle \Phi_1, Q \Phi_1  \rangle
-{g \over 3} \langle \Phi_1, \Phi_1 \star \Phi_1  \rangle .
\label{eq:ginvS23}
\end{equation}
Here, $Q$ is the BRST operator~\cite{Kato:1983im}, $\star$ is the star product of string fields, and $\Phi_1$ is the string field expanded by string Fock space states $\ket{F_i}$ of ghost number 1 with the corresponding fields $\phi_{F_i,1}$ as
\begin{equation} 
\Phi_1 = \sum_i \ket{F_i}_1\, \phi_{F_i,1}.
\label{eq:Phi1}
\end{equation}
This action is invariant under the gauge symmetry 
\begin{equation} 
\delta \Phi_1 = Q \Lambda_0 + 
g (\Phi_1 \star \Lambda_0 - \Lambda_0 \star \Phi_1)
\end{equation}
where $\Lambda_0$ is a general Grassmann even string field with ghost number $0$.
To fix this gauge symmetry, 
we take the general gauge fixing condition given by the linear equation of $\Phi_1$ as
\begin{equation}
{\rm bpz}({\cal O}^{\langle 3 \rangle} )\Phi_1=0 .
\label{eq:gaugephi1}
\end{equation}
Here ${\rm bpz}({\cal O}^{\langle 3 \rangle} )$ is an operator with ghost number $-1$ and it can be chosen arbitrarily so that the condition exactly fixes the gauge symmetry.
Note that we have represented this operator as in the form of the BPZ conjugation of 
the operator ${\cal O}^{\langle 3 \rangle} $ for later convenience. 
The superscript {\small $\langle 3 \rangle$} means that $ {\cal O}^{\langle 3 \rangle}  $ is generally supposed to operate on ghost number 3 states $\ket{F}_3$. 
Since $ {\cal O}^{\langle 3 \rangle}  $ has the ghost number $-1$, ${\cal O}^{\langle 3 \rangle}  \ket{F}_3$ has non-zero inner product only with respect to states with ghost number $1$. 
Thus, in general the operator ${\rm bpz}({\cal O}^{\langle 3 \rangle} )$ is to operate only on ghost number 1 string fields.

For each linear gauge fixing condition specified by ${\rm bpz}({\cal O}^{\langle 3 \rangle} )$, we can write the gauge fixed action as~\cite{Asano:2006hk} 
\begin{equation}
S_{\rm GF} = 
-{1 \over 2}  \sum_{n=-\infty}^{\infty} \left\langle \Phi_n, Q  \Phi_{-n+2} \right\rangle
-{g \over 3} \sum_{l+m+n=3} \left\langle  \Phi_l, \Phi_m \star \Phi_n 
\right\rangle 
+  \sum_{n=-\infty}^{\infty} 
\left\langle {\cal O}^{\langle -n+4 \rangle}  {\cal B}_{-n+4} , \Phi_{n}  \right\rangle 
\label{eq:SGF}
\end{equation}
where $\Phi_n$ and ${\cal B}_{n}$ are Grassmann odd string fields.
They are expanded by the ghost number $n$ states with the corresponding Grassmann parity $(-1)^{n-1}$ component fields. For example, $\Phi_n$ is written as the form $\Phi_n = \sum_i \ket{F_i}_n\, \phi_{F_i,n}$ with ghost number $n$ states $\ket{F_i}_n$ and the Grassmann parity $(-1)^{n-1}$ fields $\phi_{F_i,n}$.
The string fields $\Phi_n$ with $n\le 0$ and those with $n\ge 2$ respectively
represent the ghost fields (or ghost for ghost fields, etc.) and the anti-ghost fields etc.
On the other hand, the auxiliary string field ${\cal B}_{n}$ plays a role of imposing the condition 
\begin{equation}
{\rm bpz}({\cal O}^{\langle -n+4 \rangle} ) \Phi_n=0
\label{eq:gaugecondn}
\end{equation}
on $\Phi_n$.
The operator ${\cal O}^{\langle n \rangle} $ has the ghost number $-1$ and generally is to operate on states with ghost number $n$. Then, the BPZ conjugation ${\rm bpz}({\cal O}^{\langle n \rangle} )$ of ${\cal O}^{\langle n \rangle} $ operates on states with ghost number $-n+4$ 
since ${\cal O}^{\langle n \rangle} \ket{F}_{n}$ can only have non-zero inner product with a state $\ket{F'}_{-n+4}$ with ghost number $-n+4$ and 
$$
 \left\langle \ket{F'}_{-n+4}\,,\, {\cal O}^{\langle n \rangle} \ket{F}_{n}   \right\rangle
=
\left\langle (-1)^{n}{\rm bpz}({\cal O}^{\langle n \rangle} ) \ket{F'}_{-n+4}\,,\, \ket{F}_{n}   \right\rangle.
$$
We choose the operator ${\cal O}^{\langle n \rangle} $ so that the gauge symmetry 
\begin{equation}
\tilde{\delta}\Phi_n = Q\Lambda_{n-1} + 
g \sum_m ( \Phi_m \star \Lambda_{n-m} - \Lambda_{n-m} \star \Phi_m )
\label{eq:extgaugesym}
\end{equation}
of the extended action 
\begin{equation}
\tilde{S} = 
-{1 \over 2}  \sum_{n=-\infty}^{\infty} \left\langle \Phi_n, Q  \Phi_{-n+2} \right\rangle
-{g \over 3} \sum_{l+m+n=3} \left\langle  \Phi_l, \Phi_m \star \Phi_n 
\right\rangle 
\end{equation}
is completely fixed and the gauge fixed action $S_{\rm GF}$ has no gauge symmetry.
Note, however, that since the gauge transformation is non-linear for non-zero $g$, we may not be able to fix the gauge symmetry completely all over the configuration space
by using a linear gauge condition.
We specifically choose the gauge condition ${\cal O}^{\langle n \rangle} $ so that the linear part ($g=0$ part) of the gauge symmetry is completely fixed as long as the string fields satisfy $L_0\,(=\alpha'p^2+N-1 ) \ne 0$.
Note that $N$ counts the level of a state and specifies the mass of the state as $\alpha' (\mbox{mass})^2=N-1$. The condition $L_0\ne 0$ is needed since the cohomology structure of $Q$ changes for the states satisfying $L_0=0$. 
Thus, even if we choose the gauge condition appropriately for general string fields with $L_0\ne 0$, there appear residual gauge symmetry if we take $L_0 =0$.

In general, the gauge fixed action $S_{\rm GF}$ with appropriately chosen ${\cal O}^{\langle n \rangle} $ has the BRST symmetry instead of the gauge symmetry.
We can show that $S_{\rm GF}$ is invariant under the BRST transformation~\cite{Asano:2006hk, Thorn:1986qj} 
\begin{eqnarray}
\delta_B \Phi_n &=& \eta {\cal O}^{\langle n+1 \rangle}  {\cal B}_{n+1} \qquad \qquad (n>1),
\label{eq:BRS1}
\\
\delta_B \Phi_{n} &=& \eta \Big(Q \Phi_{n-1} + g \sum_{k=-\infty}^{\infty}(\Phi_{n-k} \star \Phi_{k}) \Big)\qquad (n\le 1),
\label{eq:BRS2}
\\
\delta_B {\cal B}_n &=& 0
\label{eq:BRS3}
\end{eqnarray}
with Grassmann odd parameter $\eta$ if the operators ${\cal O}^{\langle n \rangle}$ satisfy
the condition 
\begin{equation}
{\rm bpz}({\cal O}^{\langle -n+5 \rangle} ) {\cal O}^{\langle n \rangle}  {\cal B}_{n} =0.
\label{eq:Ocond}
\end{equation}
Note also that if this condition is satisfied, 
the action $S_{\rm GF}$ is invariant under another transformation 
\begin{eqnarray}
\delta'_B \Phi_n &=& 
\eta \Big(Q \Phi_{n-1} + g \sum_{k=-\infty}^{\infty}(\Phi_{n-k} \star \Phi_{k}) \Big)
 \qquad (n>1),
\\
\delta'_B \Phi_{n} &=& 
\eta {\cal O}^{\langle n+1 \rangle}  {\cal B}_{n+1} 
\qquad\qquad (n\le 1),
\\
\delta'_B {\cal B}_n &=& 0.
\end{eqnarray}

\subsection{Propagators}
Now we analyze the quadratic part of the gauge fixed action $S_{\rm GF}$ 
and derive the propagator for a given choice of ${\cal O}^{\langle n \rangle} $. 
In our discussion, we restrict the form of ${\cal O}^{\langle n \rangle} $ as
\begin{equation}
{\cal O}^{\langle n \rangle}  ={\cal O}^{\langle n \rangle} c_0b_0
\label{eq:Oc0b0}
\end{equation}
and thus 
${\cal B}_{n}$ as 
\begin{equation}
{\cal B}_{n}=c_0b_0{\cal B}_{n}.
\label{eq:Bc0b0}
\end{equation}
We need this restriction only for practical reasons: 
With this restriction, we confirm that ${\cal O}^{\langle n \rangle} $ does give conditions on the string field no more than necessary and we can analyze the propagator straightforwardly.
The most of the gauge conditions including Siegel gauge, $a$-gauges~\cite{Asano:2006hk}, and linear $b$-gauges~\cite{Kiermaier:2007jg} are within this class.
We will later give an example of possible representations of ${\cal O}^{\langle n \rangle} $ alternative to eq.(\ref{eq:Oc0b0}) which is more convenient for dealing with linear $b$-gauges.
 
For a given choice of ${\cal O}^{\langle n \rangle} $ satisfying eq.(\ref{eq:Oc0b0}), the quadratic part of the gauge fixed action $S_{\rm GF}$ is written as 
\begin{eqnarray}
S_{\rm GF}^{(2)} & = &
-{1 \over 2}  \sum_{n=-\infty}^{\infty} \left\langle \Phi_n, Q  \Phi_{-n+2} \right\rangle
+  \sum_{n=-\infty}^{\infty} 
\left\langle {\cal O}^{\langle -n+4 \rangle}  {\cal B}_{-n+4} , \Phi_{n}  \right\rangle 
\nonumber\\
&=&
-{1 \over 2}  \sum_{n=-\infty}^{\infty} 
\left\langle
(\Phi_n \;\; {\cal B}_{n+2} ), \;
\left(
\begin{array}{cc}
   Q & -{\cal O}^{\langle -n+4 \rangle}  \\
   {\rm bpz}({\cal O}^{\langle n+2 \rangle} )  & 0
  \end{array}
\right)
\left(
\begin{array}{c}
   \Phi_{-n+2} \\
   {\cal B}_{-n+4}
  \end{array}
\right)
\right\rangle.
\end{eqnarray}
If we have properly chosen ${\cal O}^{\langle n \rangle} $, the operator matrix in the second line has inverse as far as $L_0 \ne 0$ is satisfied.
The inverse matrix generally operates on the string field vector of the form $(\Psi_{-n+3},\; b_0c_0\Psi_{-n+1})^{\rm T}$ and has the form 
\begin{equation}
\left(
\begin{array}{cc}
   \Delta^{\langle -n+3 \rangle}  & A^{\langle -n+1 \rangle}  \\
   {\rm bpz}(A^{\langle n+1 \rangle} )  & 0
  \end{array}
\right).
\end{equation}
Here $\Delta^{\langle n \rangle} $ is an operator with ghost number $-1$ and 
operates on the ghost number $n$ string fields.
The operator $\Delta^{\langle n \rangle} $ is the propagator for the component fields $\phi_{F_i}$. 
Note that ${\rm bpz}(\Delta^{\langle n \rangle} )$ operates on the ghost number $4-n$ string fields and there is a relation 
\begin{equation}
{\rm bpz}(\Delta^{\langle n \rangle} ) = \Delta^{\langle 4-n \rangle} .
\end{equation}
On the other hand, $A^{\langle n \rangle} $ operates on the ghost number $n$ string fields and has ghost number $+1$.
Thus, ${\rm bpz}(A^{\langle n \rangle} )$ operates on the ghost number $2-n$ string fields.
This operator $A^{\langle n \rangle} $ (or ${\rm bpz}(A^{\langle n \rangle} )$) gives the propagator connecting $\phi_{F_i}$ and the component field of ${\cal B}_n$.

These operators are determined by the equation 
\begin{equation}
\left(
\begin{array}{cc}
   \Delta^{\langle -n+3 \rangle}  & A^{\langle -n+1 \rangle}  \\
   {\rm bpz}(A^{\langle n+1 \rangle} )  & 0
  \end{array}
\right)
\left(
\begin{array}{cc}
   Q & -{\cal O}^{\langle -n+4 \rangle}  \\
   {\rm bpz}({\cal O}^{\langle n+2 \rangle} )  & 0
  \end{array}
\right)
=
\left(
\begin{array}{cc}
   1 & 0 \\
   0 & c_0b_0
  \end{array}
\right).
\end{equation}
Note that the BPZ conjugation of this equation also holds.
From the equation, ${A}^{\langle n \rangle} $ must have the form ${A}^{\langle n \rangle} b_0c_0$ and satisfy the relation $Q A^{\langle n \rangle}=0$. 
Since the $Q$-cohomology is trivial for $L_0\ne 0$, $A^{\langle n \rangle} $ should have the form 
\begin{equation}
A^{\langle n \rangle}  = Q \tilde{A}^{\langle n \rangle} b_0c_0. 
\end{equation} 
Here the ghost number 0 operator $\tilde{A}^{\langle n \rangle} $ is determined by the equation 
\begin{equation}
{\rm bpz}({\cal O}^{\langle -n+3 \rangle} ) Q \tilde{A}^{\langle n \rangle}  b_0c_0 = b_0c_0 .
\label{eq:tildeAdef}
\end{equation} 
Since $\tilde{A}^{\langle n \rangle} $ can only be determined up to $Q \Lambda$, 
we can for example assume $[b_0,\tilde{A}^{\langle n \rangle} ]=0$. 
We define the operator $T^{\langle n \rangle} $ as
\begin{equation}
T^{\langle n \rangle}  \equiv Q {\cal O}^{\langle n \rangle}  +  {\rm bpz}({\cal O}^{\langle 3-n \rangle})  Q.
\label{eq:Tn}
\end{equation}
Then from eq.(\ref{eq:tildeAdef}) and the assumption for $\tilde{A}^{\langle n \rangle} $ given above, we have
\begin{equation}
T^{\langle n \rangle}  \tilde{A}^{\langle n \rangle}  b_0c_0 = b_0c_0 ,
\label{eq:TnAn}
\end{equation}
which determines $\tilde{A}^{\langle n \rangle} $ explicitly.
Formally, we can write $\tilde{A}^{\langle n \rangle} =(T^{\langle n \rangle}  b_0c_0  )^{-1} $.

The propagator $\Delta^{\langle n \rangle} $ can be determined by the following equations defined on ghost number $n$ string fields:
\begin{eqnarray}
&& \Delta^{\langle n+1 \rangle}  Q \; + \;Q \tilde{A}^{\langle n-1 \rangle}  \, {\rm bpz}({\cal O}^{\langle -n+4 \rangle} )  = 1,
\label{eq:propdet1}
\\
&&
Q \Delta^{\langle n \rangle}  \quad + \; {\cal O}^{\langle n+1 \rangle}  \, {\rm bpz}(\tilde{A}^{\langle -n+2 \rangle} )\, Q  = 1,
\label{eq:propdet2}
\\
&&
\Delta^{\langle n-1 \rangle}  {\cal O}^{\langle n \rangle}   = 
 {\rm bpz}({\cal O}^{\langle -n+3 \rangle} ) \Delta^{\langle n \rangle} 
=0.
\label{eq:propdet3}
\end{eqnarray} 
From these relations, we find 
\begin{equation}
\Delta^{\langle n \rangle}  Q \Delta^{\langle n \rangle}  =\Delta^{\langle n \rangle} , \qquad 
Q \Delta^{\langle n+1 \rangle}  Q =Q ,
\end{equation}
which means that the operators $Q \Delta^{\langle n \rangle} $ and $\Delta^{\langle n+1 \rangle}  Q$
are projection operators on the space of ghost number $n$ states.
Then again from the triviality of $Q$-cohomology on the state space (for $L_0\ne 0$), 
we derive 
\begin{eqnarray}
\Delta^{\langle n+1 \rangle}  Q &=& {\cal O}^{\langle n+1 \rangle}  \, {\rm bpz}(\tilde{A}^{\langle -n+2 \rangle} )\, Q,
\label{eq:propdet1'}
\\
Q \Delta^{\langle n \rangle}  &=& Q \tilde{A}^{\langle n-1 \rangle}  \, {\rm bpz}({\cal O}^{\langle -n+4 \rangle} )
\label{eq:propdet2'}
\end{eqnarray}
from eqs.(\ref{eq:propdet1}) and (\ref{eq:propdet2}).
Thus we find the general relation 
\begin{equation}
Q \Delta^{\langle n \rangle}  + \Delta^{\langle n+1 \rangle}  Q = 1 ,
\label{eq:propcomm}
\end{equation}
which, as we shall see in the next section, is the crucial property for proving the gauge independence of the 
on-shell amplitudes.
Also, $\Delta$ should satisfy 
\begin{equation}
\Delta^{\langle n \rangle}  \Delta^{\langle n+1 \rangle}  = 0 
\end{equation}
on the space of ghost number $n+1$ states.  
Note that the condition eq.(\ref{eq:Ocond}) for the BRST invariance of $S_{\rm GF}$ 
is derived from eqs.(\ref{eq:propdet3}) and (\ref{eq:propcomm}). This means that we need the condition eq.(\ref{eq:Ocond}) from the beginning in order to obtain the propagator consistently.

The explicit form of $\Delta^{\langle n \rangle} $ is determined as follows.
From eqs.(\ref{eq:propdet1'}) and (\ref{eq:propdet2'}), 
$\Delta^{\langle n \rangle} $ can be represented as 
\begin{equation}
\Delta^{\langle n \rangle}   = 
{\cal O}^{\langle n \rangle}  \, {\rm bpz}(\tilde{A}^{\langle -n+3 \rangle} ) 
Z^{\langle n+1 \rangle}  Q
\end{equation}
or
\begin{equation}
\Delta^{\langle n \rangle}  =  \tilde{A}^{\langle n-1 \rangle}  \, {\rm bpz}({\cal O}^{\langle -n+4 \rangle} )
-Q {\rm bpz}(Z^{\langle -n+5 \rangle} ) 
\end{equation}
with some ghost number $-2$ operators $Z^{\langle n \rangle} $.
Then, from the relation $ \Delta^{\langle n \rangle}  =  \Delta^{\langle n \rangle}  Q \Delta^{\langle n \rangle} $, 
we see that the propagator $\Delta^{\langle n \rangle} $ is explicitly determined as 
\begin{equation}
\Delta^{\langle n \rangle}  =  
{\cal O}^{\langle n \rangle}  \, {\rm bpz}(\tilde{A}^{\langle -n+3 \rangle} ) 
Q \tilde{A}^{\langle n-1 \rangle}  \, {\rm bpz}({\cal O}^{\langle -n+4 \rangle} ).
\label{eq:defprop1}
\end{equation}

We consider the relation between the propagators $\Delta_0^{\langle n \rangle} =b_0/L_0$ for Siegel gauge and those $\Delta^{\langle n \rangle} $ of another arbitrary linear gauge.  
Both propagators satisfy the relation (\ref{eq:propcomm}), and thus from the property of $Q$-cohomology, the difference of them can be written in the form 
\begin{equation}
\Delta^{\langle n \rangle}  - \Delta_0 = Q X^{\langle n \rangle}  - X^{\langle n+1 \rangle}  Q 
\label{eq:deffD0D}
\end{equation}
where $X^{\langle n \rangle} $ is a ghost number $-2$ operator which satisfies 
${\rm bpz}(X^{\langle n \rangle} ) = X^{\langle 5-n \rangle} $.
From eqs.(\ref{eq:propdet1'}) and (\ref{eq:propdet2'}), 
\begin{equation}
Q X^{\langle n \rangle}  Q =
 {\cal O}^{\langle n \rangle}  \, {\rm bpz}(\tilde{A}^{\langle -n+3 \rangle} ) Q 
- \Delta_0 Q,
\label{eq:DX*}
\end{equation}
and we can determine $X^{\langle n \rangle} $ up to $Q\Lambda+\Lambda'Q$ from this equation. 
For an arbitrary representative $X^{\langle n \rangle} _*$ of $X^{\langle n \rangle} $, 
we can obtain an explicit form of $\Delta^{\langle n \rangle} $ as 
\begin{equation}
\Delta^{\langle n \rangle}  = \Delta_0 + Q X_*^{\langle n \rangle}  - X_*^{\langle n+1 \rangle}  Q 
+Q(\Delta_0 X_*^{\langle n+1 \rangle}  - X_*^{\langle n \rangle}  \Delta_0 - X_*^{\langle n \rangle}  Q X_*^{\langle n+1 \rangle}   )Q.
\label{eq:defprop}
\end{equation}

To summarize, we have obtained the propagator $\Delta^{\langle n \rangle} $ explicitly as in the form of (\ref{eq:defprop1}) or (\ref{eq:defprop}) for a given choice of 
$ {\cal O}^{\langle n \rangle}  = {\cal O}^{\langle n \rangle}  c_0b_0$.  
Note that we can also proceed the similar discussion under the other representation of ${\cal O}^{\langle n \rangle} $. 
For example, consider to replace $c_0b_0$ in eq.(\ref{eq:Oc0b0}) with a projection operator $C^{\langle n-1 \rangle}  B^{\langle n \rangle} $ consisting of a ghost number $+1$ operator $C^{\langle n-1 \rangle} $ and a ghost number $-1$ operator $B^{\langle n \rangle} $.
We choose these operators to satisfy 
$B^{\langle n \rangle}  {\rm bpz}(B^{\langle 3-n \rangle})  =0$ and 
$ C^{\langle n-1\rangle}  B^{\langle n\rangle}  + 
{\rm bpz}(B^{\langle 3-n \rangle})  {\rm bpz}(C^{\langle 2-n \rangle}) =1$.
If the choice of ${\cal O}^{\langle n \rangle}  = B^{\langle n \rangle} $ gives an appropriate gauge fixing condition, then our discussion given in this subsection for $ {\cal O}^{\langle n \rangle}  = {\cal O}^{\langle n \rangle}  c_0b_0$ can be applied for the case of $ {\cal O}^{ \langle n \rangle } = {\cal O}^{\langle n \rangle}  C^{\langle n-1 \rangle}  B^{\langle n \rangle}  $ straightforwardly and the propagator is obtained similarly.

\subsection{Examples}
Now we give two examples of linear gauge conditions whose gauge fixed actions can be written in the BRST invariant form eq.(\ref{eq:SGF}).
One is the $a$-gauges corresponding to the one-parameter covariant gauges of the gauge field theory proposed in ref.\cite{Asano:2006hk}.
The other is a class of linear $b$-gauges investigated in ref.\cite{Kiermaier:2007jg} given in general by ${\cal O}^{\langle -n+4 \rangle}$ with a linear combination of $b_m$.
The Schnabl gauge~\cite{Schnabl:2005gv} and the Siegel gauge are both within this class.
For each example, we calculate the propagators explicitly.

\paragraph{$a$-gauges}
First example is the one-parameter $a$-gauges proposed in ref.\cite{Asano:2006hk}.
These gauges have one to one correspondence with the well-known covariant gauges of the gauge theory. 
They include Feynman (Siegel) gauge ($a=0$) and Landau gauge ($a=\infty$).

The gauge condition for a parameter $a\, (\ne 1)$ (including $a=\infty$) is determined by choosing the operators ${\rm bpz}\,{\cal O}$ as%
\footnote{
Here we use the slightly different (but certainly equivalent) form of the representation ${\rm bpz}\,{\cal O}$ compared to the original definition of ref.\cite{Asano:2006hk}. 
} 
\begin{eqnarray}
{\rm bpz}({\cal O}_a^{\langle n+1 \rangle} ) &=& \frac{1}{1-a}(b_0 +ab_0c_0 W_{n-1}M^{n-2}\tilde{Q}) \hspace*{3cm} (n\ge 2),
\\
{\rm bpz}({\cal O}_a^{\langle -n+4 \rangle} ) &=& b_0 (1-P_{n-2})+
\frac{1}{1-a}(b_0 P_{n-2} +ab_0c_0 \tilde{Q} M^{n-2} W_{n-1})\qquad (n\ge 2)
\end{eqnarray}
where $\tilde{Q}$ and $M$ are operators without including the ghost zero modes and are
defined from $Q = \tilde{Q} +c_0 L_0 + b_0 M$. Note that $M = -\sum_{n>0} 2n c_{-n} c_{n}$.
The operator $W_{n}$ $(n> 0)$ is defined only on the state space $\{b_0 c_0 \ket{F}_{n+1} \}$ or $\{c_0 b_0 \ket{F}_{n+2}\}$ 
($\ket{F}_{n}$ is a state with ghost number $n$) and gives the inverse operator of $M^n$ in the sense that it satisfies
$ (M^{n} W_{n}-1)$ $\times b_0 c_0 \ket{F}_{n+1} = 0$
and $( W_{n} M^{n} -1)  b_0 c_0\ket{F}_{-n+1} = 0$ 
(and $ (M^{n} W_{n} -1) c_0 b_0 \ket{F}_{n+2} = 0$ and $( W_{n} M^{n} -1)  c_0 b_0\ket{F}_{-n+2} = 0$)~\cite{Asano:2006hk}.
Explicitly, we can write 
\begin{equation}
W_n =\sum_{i=0}^{\infty}\, (-1)^i \frac{(n+i-1)!}{[(n+i)!]^2\, i!\,(n-1)!}\,
(M^-)^{n+i} M^i 
\end{equation}
where $M^- = -\sum_{n>0} \frac{1}{2n} b_{-n} b_{n}$. 
Note that $W_n$ given by the above equation can act on any state with any ghost number. This operator, however, if acted on a state not within the space $\{b_0 c_0 \ket{F}_{n+1} \}$ or $\{c_0 b_0 \ket{F}_{n+2}\}$, does not play a role of an inverse operator of  $M^{n}$ any more. 
Note also that $M^{n}$ does not have inverse originally in the whole state space.
We can only define the inverse operator of $M^n$ on the limited subspaces $\{b_0 c_0 \ket{F}_{-n+1} \}$ and $\{c_0 b_0 \ket{F}_{-n+2}\}$.
The operator $P_n$ ($n\ge 0$) is defined on the space 
$\{b_0 c_0 \ket{F}_{n+1} \}$ or $\{c_0 b_0 \ket{F}_{n+2}\}$ for $L_0\ne 0$, and is given by 
\begin{equation}
P_n=-\frac{1}{L_0} M^n \tilde{Q} W_{n+1} \tilde{Q}.
\end{equation}
Since $\tilde{Q}^2=-L_0 M$, the relations $\tilde{Q} P_n = \tilde{Q}$ and $P_n^2=P_n$ hold. 
Thus, the operator $P_n$ is a projection operator on the space $\{b_0 c_0 \ket{F}_{n+1} \}$ or $\{c_0 b_0 \ket{F}_{n+2}\}$.
In particular, the operator $1-P_n$ gives the projection onto the space of states satisfying $\tilde{Q}\ket{F}_n=0$. 

We have already shown in ref.\cite{Asano:2006hk} that the above gauges represented by ${\cal O}_a^{\langle n \rangle} $ completely fix the gauge symmetries and satisfy the condition (\ref{eq:Ocond}). Thus, we can determine the propagator $\Delta_a^{\langle n \rangle} $ by following the general procedure given in the previous subsection.
First, we calculate the operator $T_a^{\langle n \rangle}$ from eq.(\ref{eq:Tn}).
Since ${\rm bpz}{\cal O}_a^{\langle n \rangle} Q =b_0 Q $ holds for any $a$ or $n$, the result is 
\begin{equation}
T_a^{\langle n \rangle} = L_0 .
\end{equation}
Thus, from eq.(\ref{eq:TnAn}), we can determine $\tilde{A}_a^{\langle n \rangle}$ 
independently of $a$ as
\begin{equation}
\tilde{A}_a^{\langle n \rangle}  = \frac{1}{L_0},
\end{equation}
and the propagator is given explicitly by ${\cal O}_a^{\langle n \rangle}$ as
\begin{equation}
\Delta_a^{\langle n \rangle}  =  
\frac{{\cal O}_a^{\langle n \rangle} }{L_0} \,
Q \,
\frac{{\rm bpz}({\cal O}_a^{\langle -n+4 \rangle})}{L_0} .
\end{equation}

We can also represent $\Delta_a^{\langle n \rangle}$ in the form of eq.(\ref{eq:defprop}).
First, from eq.(\ref{eq:DX*}), we can choose one appropriate $X_{*\,a}$ as
\begin{eqnarray}
X_{*\,a}^{\langle n+1 \rangle}  &=& \frac{a}{1-a} \frac{1}{L_0} M^{n-2} W_{n-1} c_0b_0 
\qquad (n \ge 2)
\\
X_{*\,a}^{\langle -n+4 \rangle}  &=& 
  -\frac{a}{1-a} \frac{1}{L_0} W_{n-1} M^{n-2} b_0c_0  \qquad (n \ge 2).
\end{eqnarray}
Then, we calculate the propagator $\Delta_a^{\langle n \rangle} $ by substituting $X_{*\,a}^{\langle n \rangle} $ into 
eq.(\ref{eq:defprop}). The result is given by
\begin{equation}
\Delta_a^{\langle n \rangle}  = \frac{b_0}{L_0} 
+ Q X_{*\,a}^{\langle n \rangle}  - X_{*\,a}^{\langle n+1 \rangle}  Q 
+Q \Xi^{\langle n+1 \rangle} _a Q
\end{equation}
where $\Xi^{\langle n+1 \rangle} _a$ is the ghost number $-3$ operator which operates on 
string fields with ghost number $n+1$ and is given by 
\begin{equation}
\Xi^{\langle n+1 \rangle} _a = 
\left\{
\begin{array}{ll}
\mbox{$\displaystyle \frac{a}{(1-a)^2} \frac{1}{L_0{}^2} M^{n-2} W_{n-1} b_0$ }
& \quad (n > 2)
\\
&
\\
\mbox{$\displaystyle \frac{2a-a^2}{(1-a)^2} \frac{1}{L_0{}^2} W_{1} b_0$ }
& \quad (n = 2)
\\
&
\\
\mbox{$\displaystyle\frac{a}{(1-a)^2} \frac{1}{L_0{}^2} W_{3-n} M^{2-n} b_0$} 
& \quad (n <2)
\end{array}
\right.
\end{equation}

Note in particular that $\Delta_a^{n=2}$ includes the propagator of the gauge field theory in covariant gauges.
This can be seen by putting $\Delta_a^{n=2}$ between the fields $c_0 \alpha^\mu_{-1}\ket{p,\downarrow}$ and $c_0 \alpha^\nu_{-1}\ket{p',\downarrow}$
($\ket{\downarrow} = c^1\ket{0}$).
This is explicitly calculated as
\begin{eqnarray}
&&\left\langle
c_0 \alpha^\mu_{-1}\ket{p,\downarrow}
, \Delta_a^{\langle n=2 \rangle}  c_0 \alpha^\nu_{-1}\ket{p',\downarrow}
\right\rangle
\nonumber\\
&& \qquad \qquad =
\left\langle
c_0 \alpha^\mu_{-1}\ket{p,\downarrow}
, 
\left[
-
\left( 1-\frac{1}{(1-a)^2} \right)\frac{P_0}{L_0}
+
\frac{1}{L_0}
\right]
 c_0 \alpha^\nu_{-1}\ket{p',\downarrow}
\right\rangle
\nonumber\\
&& \qquad \qquad =
 \frac{1}{\alpha' p^2}
 \left[-
\left( 1-\frac{1}{(1-a)^2} \right)\frac{p^\mu p^\nu}{p^2} +\eta^{\mu\nu} 
  \right]
\delta(p+p').
\end{eqnarray}
We see that the result precisely coincides with  the form of the propagator of the gauge theory in the usual covariant gauge which is often parameterized by $\alpha = \frac{1}{(1-a)^2}$.

\paragraph{Linear $b$-gauges}
Next, we consider the example of linear $b$-gauges \cite{Kiermaier:2007jg}.
The gauge condition is given by  
\begin{equation}
{\rm bpz}({\cal O}_{v}^{\langle -n+4 \rangle} ) \Phi_n =0 
\end{equation}
where ${\cal O}_{v}^{\langle n \rangle}$ is made up of linear combination of $b_m$ as
\begin{equation}
{\rm bpz}({\cal O}_{v}^{\langle n=2k+1 \rangle} )= \sum_m v_m b_m
\;\;(\equiv B_v),
\qquad
{\rm bpz} ({\cal O}_{v}^{\langle n=2k \rangle} ) = {\rm bpz}(\sum_m v_m b_m)
\;\;(\equiv B^{\star}_v).
\label{eq:bgaugeOp}
\end{equation}
Note that the condition eq.(\ref{eq:Ocond}) is satisfied 
since $ ( \sum_m v_mb_m)^2 =0$.

If we assume $v_0\ne 0$, the gauge condition can be written as 
\begin{equation}
(b_0 + b^{(n)}_{\ne 0}) \Phi_{n} = 0
\end{equation}
where $b^{(2k+1)}_{\ne 0}= \sum_m (v_m/v_0) b_m$ and $b^{(2k)}_{\ne 0}= \sum_m (-1)^{m}(v_m/v_0) b_{-m}$.
By dividing $\Phi_n$ into two parts as $\Phi_n=  [b_0c_0\Phi_n] + [c_0b_0 \Phi_n]$, the gauge condition is rewritten as
\begin{equation}
\left[b_0 (c_0b_0\Phi) + b_{\ne 0} (b_0c_0\Phi) \right]
+
\left[b_{\ne 0}  (c_0b_0\Phi) \right]=0
\end{equation}
where the quantity in each bracket should vanish independently.
However, the second equation $\left[b_{\ne 0}  (c_0b_0\Phi) \right]=0$ is obtained from the first one since 
$b_{\ne 0} \left[b_0 (c_0b_0\Phi) + b_{\ne 0} (b_0c_0\Phi) \right]=0$ and
$(b_{\ne 0})^2=0$.
Thus, we conclude that the gauge condition can be rewritten as
\begin{equation}
b_0c_0 (b_0 + b^{(n)}_{\ne 0}) \Phi_{n} = 0,
\end{equation}
which means that we can replace 
${\rm bpz}({\cal O}_{v}^{\langle n \rangle} )$ of eq.(\ref{eq:bgaugeOp}) with 
$b_0c_0{\rm bpz}({\cal O}_{v}^{\langle n \rangle} )$. 
Thus we can obtain the consistent gauge fixed action as far as we choose the condition ${\cal O }^{\langle n \rangle} _{v}$ so as to fix the gauge symmetry properly.
Then, by following the procedure given in the previous subsection, the propagators are given by
\begin{eqnarray}
\Delta_v^{\langle 2k+1 \rangle} &=&
B^\star_{v} c_0b_0 \,{\rm bpz}(\tilde{A}^{\langle 2k \rangle}) \,Q
\tilde{A}^{\langle 2k+1 \rangle}\, b_0c_0 B_{v},
\label{eq:propb1a}
\\
\Delta_v^{\langle 2k \rangle} &=&
B_{v} c_0b_0\, {\rm bpz}(\tilde{A}^{\langle 2k+1 \rangle}) \,Q
\tilde{A}^{\langle 2k \rangle} \, b_0c_0 B^\star_{v}
\label{eq:propb1b}
\end{eqnarray}
where $\tilde{A}^{\langle n \rangle}$ is obtained from eq.(\ref{eq:TnAn}) 
by using 
\begin{equation}
T_v^{\langle 2k+1 \rangle} = Q B^\star_{v}c_0b_0 +b_0c_0  B^\star_{v}Q,
\qquad
T_v^{\langle 2k \rangle} = Q B_v c_0b_0 +b_0c_0  B_{v}Q
\;\;[={\rm bpz}(T_v^{\langle 2k+1 \rangle})].
\end{equation}

Alternatively, we can calculate the propagators by using the original expression eq.(\ref{eq:bgaugeOp}) of ${\cal O}_{v}^{\langle n \rangle}$ by following the procedure explained in the last part of the previous subsection.
To be precise, 
we take $C^{\langle 2k \rangle} B^{\langle 2k+1 \rangle} = 
C_v^{\star} B_v^{\star}$ and 
$C^{\langle 2k-1 \rangle}B^{\langle 2k \rangle} = C_v B_v$
where $C_v$ and its BPZ conjugation $C_v^{\star}$ are defined from the equation $\{C_v,B_v\}=1$. 
The operators $T_v^{\langle n \rangle }$ in this representation become  
$T_v^{\langle 2k+1 \rangle} = \{Q, B^\star_{v}\}\;(\equiv{\cal L}_v^\star)$ and 
$T_v^{\langle 2k \rangle} = \{Q, B_{v}\}\;(\equiv{\cal L}_v)$.
If we formally write the inverse of ${{\cal L}_v}$ as $1/{{\cal L}_v}$, 
the propagators are represented as
\begin{equation}
\Delta_v^{\langle 2k+1 \rangle} =
\frac{B^\star_{v}}{{\cal L}_v^\star }
Q
\frac{B_{v}} {{\cal L}_v},
\qquad
\Delta_v^{\langle 2k \rangle} =
\frac{B_{v}}{{\cal L}_v}
Q
\frac{B^\star_{v}}{{\cal L}_v^\star}
\label{eq:propb2}
\end{equation}
which naturally coincide with the propagators obtained in ref.\cite{Kiermaier:2007jg}. 

We have obtained two different representations, eqs.(\ref{eq:propb1a}) and (\ref{eq:propb1b}), and eq.(\ref{eq:propb2}), of the same propagators for linear $b$-gauges.
They are both represented in the form of an infinite sum of operators 
consisting of $b_m$, $c_m$, and $L_m$ which are given as the coefficients of power series expansion of $b(z)$, $c(z)$, and $T_{zz}(z)$ in general,
though the latter representation, eq.(\ref{eq:propb2}), rather conforms to 
another expansion of the fields $\tilde b( \tilde{z} )$, 
$\tilde c( \tilde{z} )$, and $\tilde T( \tilde{z} )$ based on a 
suitable choice of conformal frame $ \tilde{z} $.

\section{Consistent definition of amplitudes in general linear gauges}
We have given the procedure of obtaining the propagators for general linear gauge fixing conditions. Now we proceed to discuss the general properties of amplitudes.
In the perturbation analysis of the cubic string field theory, 
we can formally compute the amplitudes by assigning the 3-string vertices $\ket{V_3}$ and 
the propagators $\Delta$ (for the corresponding gauge we use) to the vertices and to the internal lines respectively, and  contracting them by 2-string vertices $\ket{V_2}$ and the external string states for each relevant diagram. 
Here $\ket{V_3}$ and $\ket{V_2}$ are known to be defined by 
\begin{eqnarray}
&& \langle \Phi^1, \Phi^2 \rangle = \bra{V_2}\Phi^1, \Phi^2\rangle,
\label{eq:23vertex}
\\
&&
\langle \Phi^1, \Phi^2 \star \Phi^3 \rangle = \bra{V_3}\Phi^1, \Phi^2, \Phi^3\rangle.
\end{eqnarray}
The 2-string vertex $\ket{V_{2}}_{12}\in {\cal H}_{(1)} \otimes {\cal H}_{(2)}$ is given explicitly by   
\begin{equation}
\ket{V_2}_{12} = \int \frac{d^{26}p}{(2\pi)^{26}} \,(c_0^{(1)} + c_0^{(2)}) \sum_{i,j} \ket{f_i, p}_{(1)} \, \ket{f_j, -p}_{(2)}\, g^{ji}
\label{eq:def2vertex}
\end{equation}
where $\{\ket{f_i, p}\}$ gives the basis of the state space $\{b_0c_0 \ket{F}\}$ and $g^{ji}$ is the inverse of the inner product matrix $g_{ij}=\langle \ket{f_i}, c_0 \ket{f_j} \rangle$ as $g^{ji}=(g_{ij})^{-1}$. 
Note that the relation $g_{ij}= (-1)^{|i|+1}g_{ji}$ holds.
Also, $g_{ij}=0$ for $(-1)^{|i|} \ne (-1)^{|j|}$. 
Here we assign $|i|=0$ (or $1$) for a Grassmann even (or odd) state $\ket{f_i}$.
Note also that the whole string Hilbert space is represented as 
${\cal H} = \{ \ket{f_i}\} \oplus \{c_0 \ket{f_i}\}$. 
For an arbitrary operator ${\cal O}$, the 2-string vertex $\ket{V_2}_{12}$ defined by eq.(\ref{eq:def2vertex}) has the property
\begin{equation}
({\cal O}^{(1)} -{\rm bpz}{\cal O}^{(2)}) \ket{V_2}_{12} =0.
\label{eq:prop2ver}
\end{equation}

Thus, in principle, the property eq.(\ref{eq:23vertex}) should hold consistently with the above explicit definition of 2-string vertex. 
However, we face an ambiguity of sign when adapting this relation eq.(\ref{eq:23vertex}) to calculate  the inner products. 
We can see the problem clearly when we contract $\ket{V_2}_{12}$ by itself by using eq.(\ref{eq:def2vertex}): From the naive calculation, the result seems to vanish because $(c_0^{(1)} + c_0^{(2)})^2 =0$, though it should not vanish and has to give the trace of the Hilbert space.
The problem occurs since there is an ambiguity in calculating 
the contraction of an element of the tensor product of two spaces, e.g., $\ket{f_1}_{(1)} \ket{f_2}_{(2)} $, with another element 
$\ket{f'_1}_{(1)} \ket{f'_2}_{(2)}$.
To see the problem explicitly, we consider to calculate ${}_{(1)}\bra{f'_1} {}_{(2)}\bra{f'_2} \ket{f_1}_{(1)} \ket{f_2}_{(2)}$.
We have different results depending on the order of calculation.
For example, if we first exchange the states ${}_{(1)}\bra{f'_1}$ and ${}_{(2)}\bra{f'_2}$ as $ {}_{(1)}\bra{f'_1} {}_{(2)}\bra{f'_2}=(-1)^{|f'_1||f'_2|} 
{}_{(2)}\bra{f'_2} {}_{(1)}\bra{f'_1} $, then the result becomes 
$
(-1)^{|f'_1||f'_2|} \bra{f'_1} \ket{f_1} \bra{f'_2}\ket{f_2}  
$.
On the other hand, if we first exchange the states 
${}_{(2)}\bra{f'_2}$ and $\ket{f_1}_{(1)}$, we have the different result 
$
(-1)^{|f'_2||f_1|} \bra{f'_1} \ket{f_1} \bra{f'_2}\ket{f_2} 
$ in general.
We have a similar sort of ambiguities for calculating the inner product using the 3-string vertex $\ket{V_3}_{123}$.

These ambiguities originally attributed to the property of inner products of string Fock space: The inner product of states with total ghost number 3 gives the real number as $\bra{0} c_{-1} c_0 c_1 \ket{0}=1$ since we have set the ghost number of the SL$(2,R)$ invariant vacuum $\ket{0}$ to be zero.
On the other hand, if we assign the ghost number $-\frac{3}{2}$ to the state $\ket{0}$ as in the original definition \cite{Witten:1985cc, Kato:1983im} respecting the proper definition of ghost number currents, such a problem might not occur. In that case, however, we instead have to determine the commutation properties of states with half-integer Grassmann property and we also have to assign the ghost number $\frac{3}{2}$ to the star product~$\star$.  

In the following argument, we choose to take the former definition of ghost number, {\it i.e.,} we assign $\ket{0}$ the ghost number $0$.
In order to avoid the ambiguities, we perform the calculation of amplitudes without using the 3-string vertices and with the minimum use of 2-string vertices. 
When we use the 2-string vertices, we will assign extra parameters with appropriate Grassmann parities to avoid the ambiguities of calculation. 
We first give the prescription of obtaining the amplitudes by a rather a priori argument. 
We will see the consistency of the definition of amplitudes through the argument of the gauge invariance of tree amplitudes in section~4.

\subsection{Tree amplitudes}
To calculate the $n$-point tree amplitudes for $n$ external fields $\phi^a_{F^a}$ ($a=1,\cdots,n$), we first provide the external string fields 
$\Phi^a=\ket{F^a}\, \phi^a_{F^a} $ and assign the propagator $\Delta$ to each internal line. 
Then we connect these string fields $\Phi^a$ and propagators $\Delta$ by using the star-products appropriately.
The amplitudes are given as the coefficient of the product of fields $\phi^1_{F^1}\cdots \phi^n_{F^n}$ after summing up all the contributions of possible diagrams.
Note that in general the amplitude vanishes if the total Grassmann parity of external fields is odd, {\it i.e.,} $(-1)^{\sum_a|\phi^a_{F^a}|}=-1$.

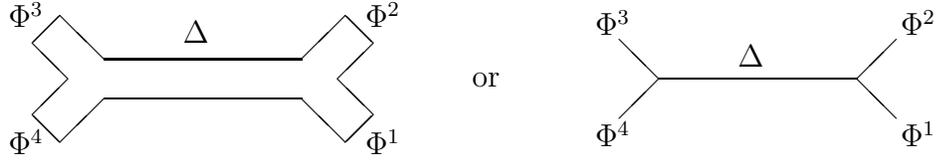
\begin{figure}[htbp]
\begin{center}
\begin{picture}(300,70)(25,0)
 \setlength{\unitlength}{1.5pt} 
\put(40,20){\line(1,0){50}}
\put(40,30){\line(1,0){50}}
\put(40,20){\line(-1,-1){11}}
\put(31,25){\line(-1,-1){9}}
\put(40,30){\line(-1,1){11}}
\put(31,25){\line(-1,1){9}}
\put(90,30){\line(1,1){11}}
\put(99,25){\line(1,1){9}}
\put(90,20){\line(1,-1){11}}
\put(99,25){\line(1,-1){9}}
\put(101,9){\line(1,1){7}}
\put(101,41){\line(1,-1){7}}
\put(29,9){\line(-1,1){7}}
\put(29,41){\line(-1,-1){7}}
\put(60,34){$\Delta$}
\put(106,6){$\Phi^1$}
\put(106,38){$\Phi^2$}
\put(16,38){$\Phi^3$}
\put(16,6){$\Phi^4$}
\put(133,23){or}
\put(180,25){\line(1,0){50}}
\put(180,25){\line(-1,-1){10}}
\put(180,25){\line(-1,1){10}}
\put(230,25){\line(1,-1){10}}
\put(230,25){\line(1,1){10}}
\put(200,28){$\Delta$}
\put(241,8){$\Phi^1$}
\put(241,36){$\Phi^2$}
\put(164,36){$\Phi^3$}
\put(164,8){$\Phi^4$}
\end{picture}
\caption{{\bf 4-point tree diagram $G_{1234}$}}
\label{fig1}
\end{center}
\end{figure}

For example, for the 4-point diagram $G_{1234}$ given in the Fig.\ref{fig1}, we have 
\begin{equation}
\langle \Phi^{1} \star \Phi^{2}, \Delta (\Phi^{3} \star \Phi^{4})  \rangle
= \phi^1_{F^1}\cdots \phi^4_{F^4}\: {A}^{G_{1234}}_{\phi^1_{F^1}\cdots \phi^4_{F^4}}
\label{eq:4pointtree}
\end{equation}
and $A^{G_{1234}}_{\phi^1_{F^1}\cdots \phi^4_{F^4}}$ gives the contribution to the 4-point tree amplitude from this diagram.
Here we have abbreviated the superscript of $\Delta^{\langle n \rangle} $ representing the ghost number of the string field on which the operator acts.
In eq.(\ref{eq:4pointtree}), $n$ is given by the sum of the ghost number of $\Phi_3$ and $\Phi_4$.
We can also represent eq.(\ref{eq:4pointtree}) as 
$\langle \Delta(\Phi^{1} \star \Phi^{2}), \Phi^{3} \star \Phi^{4} \rangle$
since ${\rm bpz}(\Delta)=\Delta$ and $\Phi^{1} \star \Phi^{2}$ (or $\Phi^{3} \star \Phi^{4}$) is even. 
In general, $\Delta$ operates only on even string fields in the calculation of any diagram since all external string fields $\Phi^a$ are odd and $\Delta$ operates on the star product of two odd string fields $\Psi\star \Psi'$ 
where $\Psi$ or $\Psi'$ generally consists of $n$ $\Phi^a$'s and $n-1$ $\Delta$'s.
Thus, the calculation of any tree diagram is consistently performed by the above procedure. 
The amplitude is obtained by summing up the contributions of all the relevant 
diagrams. For example, the 4-point tree amplitude $A_{\phi_{F^1}\cdots \phi_{F^4}}$ can be calculated as
\begin{equation}
\frac{1}{2}
\sum_{\{a_1,\cdots,a_4\}}
\langle \Phi^{a_1} \star \Phi^{a_2}, \Delta (\Phi^{a_3} \star \Phi^{a_4})  \rangle
= 
\phi^1_{F^1}\cdots \phi^4_{F^4} A_{\phi^1_{F^1}\cdots \phi^4_{F^4}}
\label{eq:4pointtreeamp}
\end{equation}
where the sum is over all permutations of $\{a_1,\cdots,a_4\}$ and 
the $\frac{1}{2}$ in front of the sum is necessary since two graphs $G_{a_1a_2a_3a_4}$ and $G_{a_3a_4a_1a_2}$ are indistinguishable.

\subsection{Loop amplitudes}
For $k$-loop amplitudes, we perform the calculation by reducing each graph to a tree graph. 
We use at least $k$ 2-string vertices $\ket{V_2}$ to calculate the contribution of each $k$-loop diagram.
In the following, we mainly consider the 1-loop amplitudes and explain the method of calculation in detail.

We first give a consistent method to calculate the contribution of general 1-loop diagrams and then discuss the consistency.
\begin{figure}[htbp]
\begin{center}
\begin{picture}(400,140)(0,0)
\put(-10,120){(a)}
\put(130,120){(b)}
\put(0,50){\includegraphics[height=80pt]{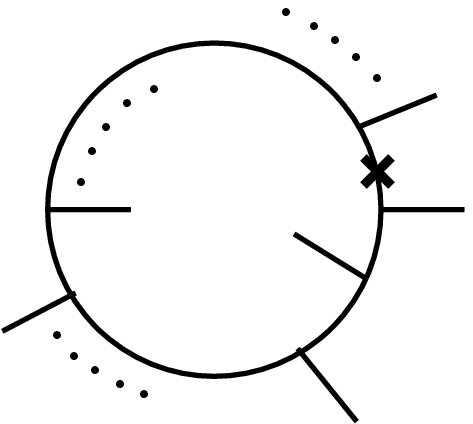}}
\put(190,0){\includegraphics[height=120pt]{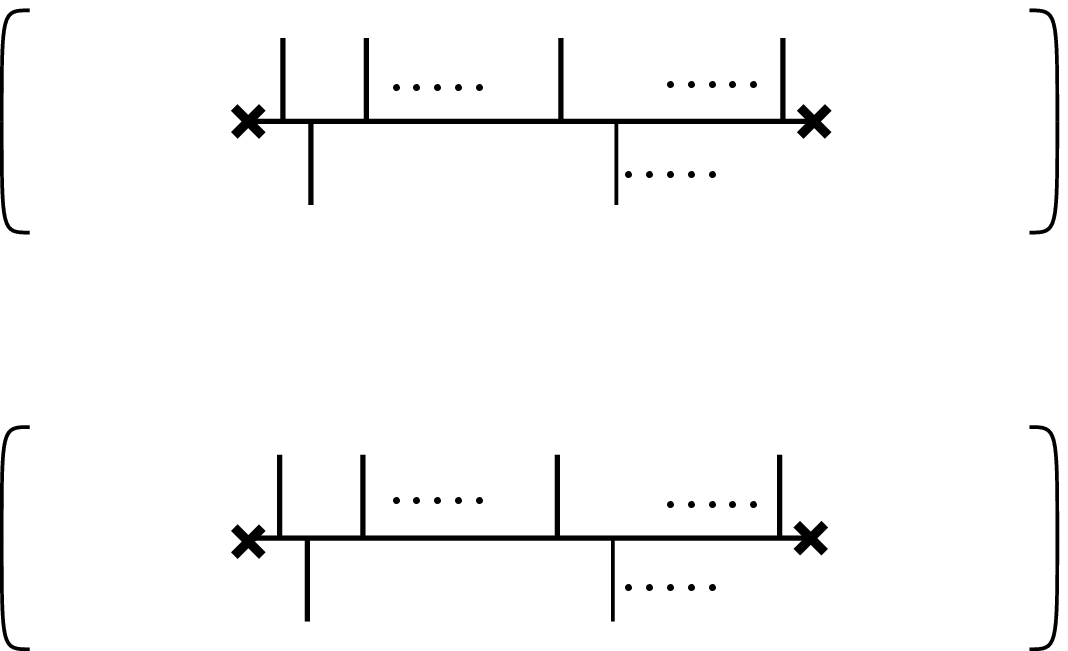}}
\put(55,91){$\alpha$}
\put(0,75){$\beta$}
\put(160,98){$\displaystyle\sum_I$}
\put(191,98){\small$\Delta\ket{V_I}\eta_{I}^{\alpha}$}
\put(346,98){\small$\ket{V'_I}\tilde{\eta}_{I}^{\alpha}$}
\put(150,60){$=$}
\put(160,18){$\displaystyle\sum_I$}
\put(195,18){\small$\ket{V'_I}\tilde{\eta}_{I}^{\alpha}$}
\put(346,18){\small$\Delta\ket{V_I}\eta_{I}^{\alpha}$}
\end{picture}
\caption{{\bf calculation of 1-loop diagrams (1):~} (a)~For a given 1-loop diagram $G$, we choose an internal line $\alpha$ so that we obtain a tree graph $G_{(\alpha)}$ by cutting the line. (b)~We cut $\alpha$ and assign $\Delta\ket{V_{I}}\eta_{I}^{\alpha}$ and $\ket{V'_{I}}\tilde{\eta}_{I}^{\alpha}$ to the two end points. We can exchange the assignment of two states as is ensured in eq.(\ref{eq:exchangeVV}).}
\label{fig:fig2}
\end{center}
\end{figure}
For a given 1-loop diagram, we `cut' an internal line $\alpha$ and assign the Grassmann odd states $\Delta\ket{V_{I}}\eta_{I}^{\alpha}$ and $\ket{V'_{I}}\tilde{\eta}_{I}^{\alpha}$ to the two end points. We choose $\alpha$ so as to obtain one tree graph if we cut the line. (See Fig.\ref{fig:fig2}.) The states $\ket{V_{I}}$ and $\ket{V'_{I}}$ are defined from the 2-string vertex $\ket{V_2}_{12}$ as
\begin{equation}
\ket{V_2}_{12} = \sum_{I} \ket{V_{I}}_{(1)} \otimes \ket{V'_{I}}_{(2)}
\;\left(=  -\sum_{I} \ket{V'_{I}}_{(1)} \otimes \ket{V_{I}}_{(2)} \right).
\label{eq:2verII'}
\end{equation}
Explicit forms of $\ket{V_{I}}$ and $\ket{V'_{I}}$ can be read from eq.(\ref{eq:def2vertex}). 
The parameters $\eta_{I}^{\alpha}$ and $\tilde{\eta}_{I}^{\alpha}$ both have Grassmann parity $(-1)^{|V_I|}(=(-1)^{|V'_I|+1})$. We only need these parameters in order that we can give the correct sign contribution to the amplitudes.
If they are placed next to each other in the order of $\eta_{I}^{\alpha} \tilde{\eta}_{I}^{\alpha}$, we can replace them to 1 as
$\eta_{I}^{\alpha} \tilde{\eta}_{I}^{\alpha} \,(\,= \tilde{\eta}_{I}^{\alpha}{\eta}_{I}^{\alpha}(-1)^{|V_I|} )\rightarrow 1$ for each $I$.
We can exchange the assignment of $\Delta\ket{V_{I}}\eta_{I}^{\alpha}$ and $\ket{V'_{I}}\tilde{\eta}_{I}^{\alpha}$ on the two ends since 
\begin{eqnarray}
\sum_I \left( \Delta^{(1)}\ket{V_{I}}\eta_{I}^{\alpha} \right)_{(1)}
\Big(\ket{V'_{I}}\tilde{\eta}_{I}^{\alpha}\Big)_{(2)}&=&
\sum_I \Delta^{(2)} \ket{V_{I}}_{(1)}
\ket{V'_{I}}_{(2)} \eta_{I}^{\alpha}\tilde{\eta}_{I}^{\alpha}
\nonumber\\
&=& 
-\sum_I \Delta^{(2)} \ket{V'_{I}}_{(1)}
\ket{V_{I}}_{(2)} \eta_{I}^{\alpha}\tilde{\eta}_{I}^{\alpha}
\nonumber\\
&=& 
\sum_I  \ket{V'_{I}}_{(1)}
\Delta^{(2)}\ket{V_{I}}_{(2)} \eta_{I}^{\alpha}\tilde{\eta}_{I}^{\alpha}(-1)^{|V_I|}
\nonumber\\
&=& 
\sum_I 
\Big(\ket{V'_{I}}\tilde{\eta}_{I}^{\alpha}\Big)_{(1)}
\left( \Delta^{(2)}\ket{V_{I}}\eta_{I}^{\alpha} \right)_{(2)}.
\label{eq:exchangeVV}
\end{eqnarray}
Here we have used the relation (\ref{eq:prop2ver}) and 
the replacement $\eta_{I}^{\alpha} \tilde{\eta}_{I}^{\alpha}\leftrightarrow 1$ properly.

Then we calculate the amplitude by treating the $n$-point 1-loop diagram (with a cut on $\alpha$) as an $(n+2)$-point tree diagram which has external string fields $\Phi^a$ ($a=1,\cdots,n$), $\Delta\ket{V_{I}}\eta_{I}^{\alpha}$ and $\ket{V'_{I}}\tilde{\eta}_{I}^{\alpha}$. 
After taking the sum over $I$ and removing $\eta_{I}^{\alpha}\tilde{\eta}_{I}^{\alpha} $, we obtain the amplitude as the coefficient of $\phi^1_{F^1}\cdots \phi^n_{F^n}$. 
To conclude, the contribution of a given 1-loop $n$-point diagram $G$ is calculated as
\begin{eqnarray}
&& {\cal F}^{G}_{\rm 1-loop}(\Phi^1,\cdots,\Phi^n)
\nonumber\\
&& = \sum_I {\cal F}^{G_{(\alpha)}}_{\rm tree}(\Phi^1,\cdots,\Phi^n, 
\Delta\ket{V_{I}}\eta_{I}^{\alpha}, \ket{V'_{I}}\tilde{\eta}_{I}^{\alpha})
\nonumber\\
&&= \phi^1_{F^1}\cdots \phi^n_{F^n} \: \mbox{\cal A}^{G}_{\phi^1_{F^1}\cdots \phi^n_{F^n}}
\end{eqnarray}
where $G_{(\alpha)}$ denotes the $(n+2)$-point tree graph obtained by cutting the line $\alpha$ of $G$.

For the consistency of this definition of 1-loop amplitudes, we have to check whether the result remains unchanged if we cut the graph on a different internal line $\beta$.
For this purpose, we first introduce the unit operator ${\bf 1}$ which operates on a state $\ket{F}$ as ${\bf 1} \ket{F} =\ket{F}$.
The explicit form of ${\bf 1}$ is given by
\begin{eqnarray}
{\bf 1} &=& \ket{f_i}g^{ji} \bra{{\rm bpz}(c_0 \ket{f_j})}
+c_0 \ket{f_i}g^{ij} \bra{{\rm bpz}(\ket{f_j})}
\\
&=&
\sum_I \ket{V'_I} \:\bra{{\rm bpz}(\ket{V_I})}\: (-1)^{|V_I|}
\end{eqnarray}
where we have used the same notation as for the 2-string vertex $\ket{V_{2}}$.
Note that there appear extra sign contributions in ${\bf 1}$ compared to the 2-string vertex unlike the naive expectation.
This sign difference between ${\bf 1}$ and $\ket{V_{2}}$ is also related to the 
sign ambiguity caused by the property of inner products.
Furthermore, since the Grassmann parity of the operator ${\bf 1}$ is odd 
(with total ghost number three) as in the case of $\ket{V_{2}}$, we must be careful about the possible sign ambiguity when we insert ${\bf 1}$ into a bracket.
In particular, if we consider brackets of string fields consisting of states and the component fields, we should insert ${\bf 1}$ after removing all the component fields 
in order to avoid the sign ambiguities. 
For example, consider the following bracket 
\begin{equation}
\langle \ket{F} \phi^{F}, \ket{G}\phi^{G}\rangle.
\end{equation} 
We can assume $(-1)^{|\mbox{\scriptsize $\ket{F}$}|+|\mbox{\scriptsize $\ket{G}$}|}=-1$ since otherwise the equation vanishes.
Also, we can assume $(-1)^{|\phi^F|+|\phi^G|}=1$ since any string amplitudes  
with $(-1)^{\sum \phi^F_{a}}=-1$ should vanish.
Then the insertion of ${\bf 1}$ is performed consistently as
\begin{eqnarray}
\langle \ket{F} \phi^{F}, \ket{G}\phi^{G}\rangle 
&=& \phi^{F} \phi^{G} (-1)^{\mbox{\scriptsize $|\ket{G}||\phi_G|$}} \langle \ket{F} , {\bf 1}\ket{G}\rangle 
\nonumber\\
&=& \phi^{F} \phi^{G} (-1)^{\mbox{\scriptsize $|\ket{G}||\phi_G|$}} 
\sum_{I}(-1)^{V_I}
\langle \ket{F} , \ket{V_I'}
\rangle\,
\langle
\ket{V_I},
 \ket{G}\rangle .
\label{eq:1apply}
\end{eqnarray}

\begin{figure}[thbp]
\begin{center}
\begin{picture}(450,280)(0,0)
\put(0,185){\includegraphics[height=80pt]{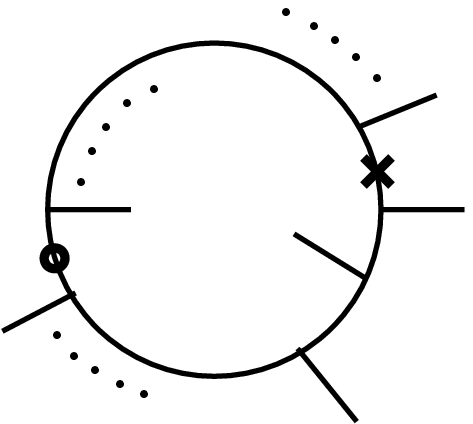}}
\put(166,0){\includegraphics[height=269pt]{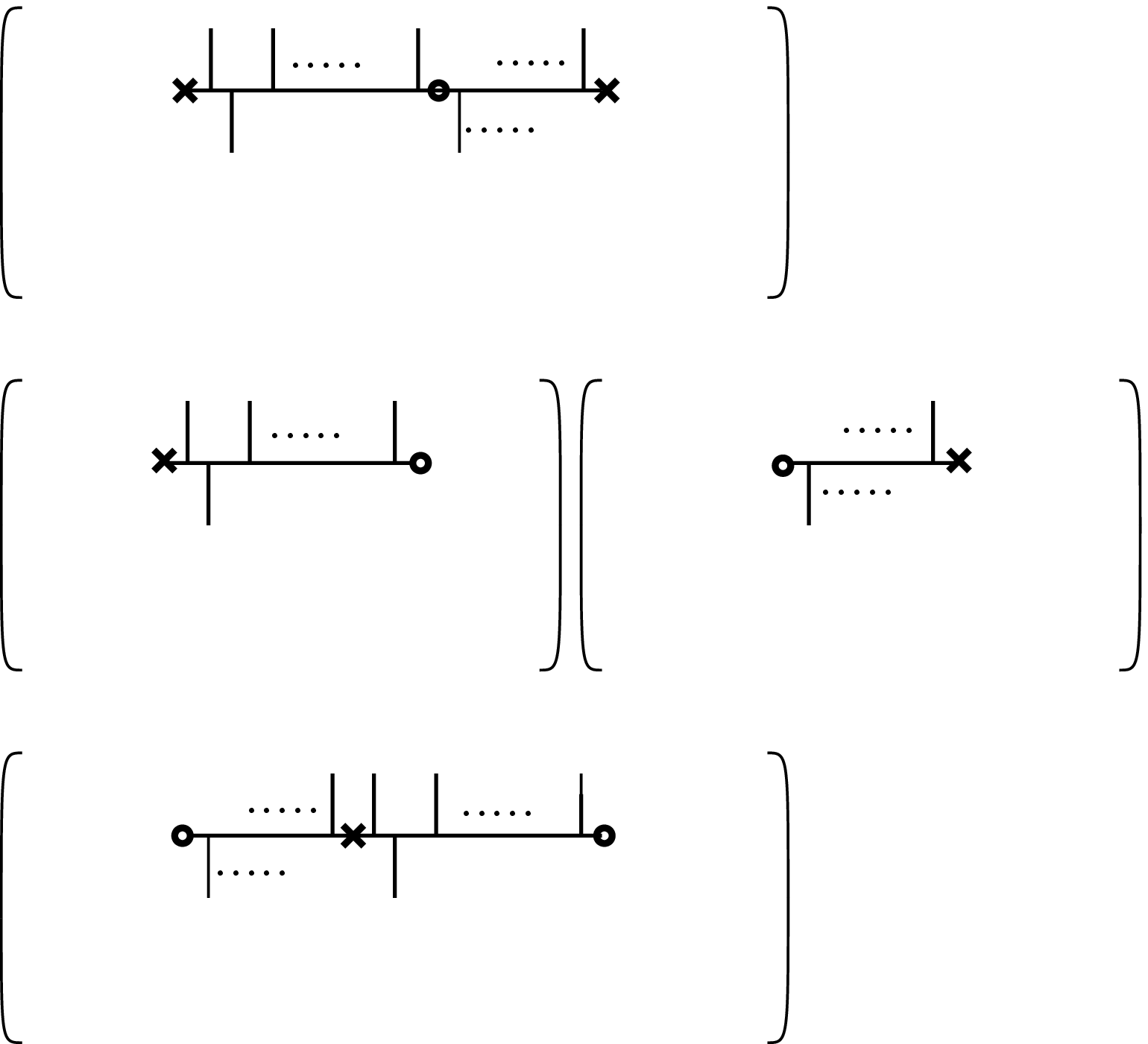}}
\put(50,226){\small$\Delta_{\alpha}$}
\put(-8,210){\small$\Delta_{\beta}$}
\put(105,225){$\Rightarrow$}
\put(145,231){$\displaystyle\sum_I$}
\put(176,244){\small$\ket{V'_I}\tilde{\eta}_{I}^{\alpha}$}
\put(328,244){\small$\Delta\ket{V_I}{\eta}_{I}^{\alpha}$}
\put(240,205){ $\displaystyle \Big\langle \Psi_A^I,  \Delta_{\beta} \Psi_B^I  \Big\rangle$}
\put(105,134){$\Leftrightarrow$}
\put(128,134){$\displaystyle\sum_I\sum_J$}
\put(173,147){\small$\ket{V'_I}\tilde{\eta}_{I}^{\alpha}$}
\put(278,147){\small$\ket{V'_J}\tilde{\eta}_{J}^{\beta}$}
\put(322,147){\small$\Delta\ket{V_J}\eta_{J}^{\beta}$}
\put(418,147){\small$\Delta\ket{V_I}\eta_{I}^{\alpha}$}
\put(200,110){$\displaystyle \left\langle \Psi_A^I,  \Big(\ket{V'_{J}} \tilde{\eta}_{J}^{\beta}\Big)
\right\rangle$}
\put(340,110){$\left\langle \Big(\Delta \ket{V_{J}}\eta_{J}^{\beta}\Big) ,  \Psi_B^I  \right\rangle 
$}
\put(105,38){$\Leftrightarrow$}
\put(145,38){$\displaystyle\sum_J$}
\put(174,50){\small$\ket{V'_J}\tilde{\eta}_{J}^{\beta}$}
\put(328,50){\small$\Delta\ket{V_J}\eta_{J}^{\beta}$}
\put(240,16){$\displaystyle \Big\langle \Psi_A^{'J},\Delta_{\beta}
\Psi_B^{'J}  \Big\rangle$}
\end{picture}
\caption{{\bf calculation of 1-loop diagrams (2):~}
A 1-loop diagram $G$ can be consistently calculated from either of the tree graphs $G_{(\alpha)}$ or $G_{(\beta)}$ obtained by cutting an internal line $\alpha$ or $\beta$ of $G$.
}
\label{fig:fig3}
\end{center}
\end{figure}

For an $n$-point diagram $G$, we have the result of calculation 
$\sum_I {\cal F}^{G_{(\alpha)}}_{\rm tree}
(\Phi^a, 
\Delta\ket{V_{I}}\eta_{I}^{\alpha}, \ket{V'_{I}}\tilde{\eta}_{I}^{\alpha})$ 
given by cutting one particular internal line $\alpha$.
Then we consider to insert the unit operator ${\bf 1}$ in front of the propagator $\Delta$ on another internal line $\beta$. (See Fig.\ref{fig:fig3}.)
In general, this $\sum_I {\cal F}^{G_{(\alpha)}}_{\rm tree}$ can be deformed to have the form 
\begin{equation}
\sum_I {\cal F}^{G_{(\alpha)}}_{\rm tree}(\Phi^a, 
\Delta\ket{V_{I}}\eta_{I}^{\alpha}, \ket{V'_{I}}\tilde{\eta}_{I}^{\alpha}) 
=
\sum_I \langle \Psi_A^I,  \Delta_{\beta} \Psi_B^I  \rangle .
\end{equation} 
Here $\Delta_{\beta}$ is the particular propagator for an arbitrary internal line $\beta$ other than $\alpha$.
Both $\Psi_A$ and $\Psi_B$ are Grassmann even string fields 
consisting in general of $k$ external string fields ($\Phi_a$'s), $k-1$ propagators (excluding those for $\alpha$ nor $\beta$), and one of the fields 
$\Delta\ket{V_{I}}\eta_{I}^{\alpha}$ or $\ket{V'_{I}}\tilde{\eta}_{I}^{\alpha}$.
We fix that $\ket{V'_{I}}\tilde{\eta}_{I}^{\alpha}$ and $\Delta\ket{V_{I}}\eta_{I}^{\alpha}$ are respectively set within  $\Psi_A$ and $\Psi_B$.
We insert {\bf 1} in front of $\Delta_{\beta}$ after removing the fields   
$\psi_A \tilde{\eta}_I^\alpha$ and $\psi_B \eta_{I}^\alpha$ from $\Psi_A$ and $\Psi_B$ as we did in eq.(\ref{eq:1apply}). 
Here $\psi_A$ (or $\psi_B$) is the product of fields $\phi^{i_1}_{F^{i_1}} \cdots \phi^{i_k}_{F^{i_k}}$ included in $\Psi_A$ (or $\Psi_B$). 
Finally, the result can be written as
\begin{equation}
\sum_{I,J} \left\langle \Psi_A^I,  \Big(\ket{V'_{J}} \tilde{\eta}_{J}^{\beta}\Big)
\right\rangle\,
\left\langle \Big(\Delta \ket{V_{J}}\eta_{J}^{\beta}\Big) ,  \Psi_B^I  \right\rangle .
\label{eq:AB1}
\end{equation}
Note that each bracket $\langle \Psi_A^I,  (\ket{V'_{J}} \tilde{\eta}_{J}^{\beta})\rangle$ or $\langle (\Delta \ket{V_{J}}\eta_{J}^{\beta}) ,  \Psi_B^I  \rangle$ does not vanish only if the total Grassmann parity of the fields containing in the bracket is even, {\it i.e.,} $(-1)^{|\psi_A|+ |\tilde{\eta}_I^{\alpha}| +|\tilde{\eta}_J^{\beta}|} =1 $ or $(-1)^{|\psi_B|+ |{\eta}_I^{\alpha}| +|{\eta}_J^{\beta}|} =1 $. 
Thus we do not have to be worry about the sign ambiguity problem for the expression eq.(\ref{eq:AB1}).

We then deform each bracket $\langle \Psi_A^I,  (\ket{V'_{J}} \tilde{\eta}_{J}^{\beta})\rangle$ or $\langle (\Delta \ket{V_{J}}\eta_{J}^{\beta}) ,  \Psi_B^I  \rangle$ by using BPZ conjugation operation appropriately so that eq.(\ref{eq:AB1}) becomes 
\begin{equation}
\sum_{I,J} \left\langle \Psi_A^{'J},  \Big(\ket{V'_{I}} \tilde{\eta}_{I}^{\alpha}\Big)
\right\rangle\,
\left\langle \Big(\Delta \ket{V_{I}}\eta_{I}^{\alpha}\Big) ,  \Psi_B^{'J}  \right\rangle 
=
\sum_J \langle \Psi_A^{'J},  \Delta_{\beta} \Psi_B^{'J}  \rangle
\end{equation}
where $\Psi_A^{'J}$ and $\Psi_B^{'J}$ are Grassmann even and include 
$\ket{V'_{J}}\tilde{\eta}_{J}^{\beta}$ and $\Delta\ket{V_{J}}\eta_{J}^{\beta}$ respectively.
This is equivalent to $\sum_J {\cal F}^{G_{(\beta)}}_{\rm tree}
(\Phi^a, 
\Delta\ket{V_{J}}\eta_{J}^{\beta}, \ket{V'_{J}}\tilde{\eta}_{J}^{\beta})$ which is obtained by cutting the internal line $\beta$ from the beginning. 
Thus, we have shown that 
\begin{equation}
\sum_I {\cal F}^{G_{(\alpha)}}_{\rm tree}
(\Phi^a, 
\Delta\ket{V_{I}}\eta_{I}^{\alpha}, \ket{V'_{I}}\tilde{\eta}_{I}^{\alpha})
=
\sum_I {\cal F}^{G_{(\beta)}}_{\rm tree}
(\Phi^a, 
\Delta\ket{V_{I}}\eta_{I}^{\beta}, \ket{V'_{I}}\tilde{\eta}_{I}^{\beta})
\end{equation}
for any 1-loop diagram $G$ and for its arbitrary internal lines $\alpha$ and $\beta$.
This concludes that our method of calculation of 1-loop amplitudes is defined consistently.

In general, $k$-loop amplitudes can be calculated similarly. For a given $k$-loop $n$-point diagram $G$, we make an $(n+2k)$-point tree graph by cutting $k$ internal lines $\alpha_i$ ($i=1,\cdots,k$) of $G$ and assign the states $\Delta\ket{V_{I_i}}\eta_{I_i}^{\alpha_i}$ and $\ket{V'_{I_i}}\tilde{\eta}_{I_i}^{\alpha_i}$ to each of the end points of the cut $\alpha_i$.
Then the contribution of the diagram $G$ to the amplitudes can be obtained after replacing $\eta_{I_i}^{\alpha_i}\tilde{\eta}_{I_i}^{\alpha_i}\rightarrow 1$ from
\begin{equation}
{\cal F}^{G}_{ \mbox{\scriptsize\rm $k$-loop} }(\{\Phi^a\}) =
{\cal F}^{G_{(\alpha_1,\cdots,\alpha_n)}}_{\rm tree}
\left(\{\Phi^a\}, \{\Delta\ket{V_{I_i}}\eta_{I_i}^{\alpha_i}\}, \{\ket{V'_{I_i}}\tilde{\eta}_{I_i}^{\alpha_i}\}\right) 
\end{equation}
as a coefficient of the products of the external fields $\phi^1_{F^1} \cdots \phi^n_{F^n}$.

\subsection{Examples}
We give some examples of calculation of amplitudes.
For tree amplitudes, we have already given the simplest example of the calculation of 4-point amplitude in eqs.(\ref{eq:4pointtree}) and (\ref{eq:4pointtreeamp}). 
Here we give a few examples of 1-loop amplitudes. 

\paragraph{The 1-loop 1-point amplitude}
There is only one kind of graph $G_{1}$ which is relevant to the 1-loop 1-point amplitude. 
Following the argument given in the previous subsection, we have 
\begin{eqnarray}
{\cal F}^{G_{1}}(\Phi^1)
&=& \sum_I \langle \Phi^1, \Delta \ket{V_I} \eta_I , \ket{V'_I} \tilde{\eta}_I    \rangle
\nonumber\\
&=& \phi_{F^1}(p_1) \left( \sum_I \int\! \frac{d^{26}p}{(2\pi)^{26}}\, \Big\langle \ket{F^1, p_1}, \,
\Delta \ket{V_I,p} \star \ket{V'_I, -p }  \Big\rangle\right)
\end{eqnarray}
where in the second line we have expressed the momentum explicitly.
Thus, the amplitude is given by 
\begin{equation}
\mbox{\cal A}^{G_1}_{\phi_{F^1}}(p_1)= 
\sum_I \int\! \frac{d^{26}p}{(2\pi)^{26}}\, \Big\langle \ket{F^1, p_1}, \,
\Delta \ket{V_I,p} \star \ket{V'_I, -p }  \Big\rangle .
\end{equation}
Note that the amplitude vanishes for $p_1 \ne 0$ from the conservation of momentum. 
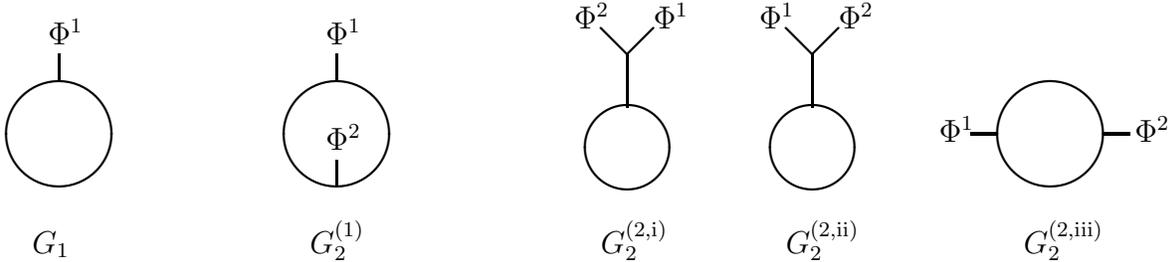
\begin{figure}[thbp]
\begin{center}
\begin{picture}(450,115)(0,-25)
\thicklines
\put(15,-20){{$G_1$}}
\put(25,25){\circle{40}}
\put(25,45){\line(0,1){10}}
\put(21,59){$\Phi^1$}
\put(120,-20){{$G_2^{\rm (1)}$}}
\put(130,25){\circle{40}}
\put(130,45){\line(0,1){10}}
\put(130,5){\line(0,1){10}}
\put(126,59){$\Phi^1$}
\put(126,19){$\Phi^2$}
\put(230,-20){{$G_2^{\rm (2,i)}$}}
\put(240,20){\circle{30}}
\put(240,35){\line(0,1){20}}
\put(240,55){\line(1,1){10}}
\put(240,55){\line(-1,1){10}}
\put(250,65){$\Phi^1$}
\put(220,65){$\Phi^2$}
\put(300,-20){{$G_2^{\rm (2,ii)}$}}
\put(310,20){\circle{30}}
\put(310,35){\line(0,1){20}}
\put(310,55){\line(1,1){10}}
\put(310,55){\line(-1,1){10}}
\put(290,65){$\Phi^1$}
\put(320,65){$\Phi^2$}
\put(390,-20){{$G_2^{\rm (2,iii)}$}}
\put(400,25){\circle{40}}
\put(380,25){\line(-1,0){10}}
\put(420,25){\line(1,0){10}}
\put(358,22){$\Phi^1$}
\put(432,22){$\Phi^2$}
\end{picture}
\caption{{\bf simple 1-loop diagrams}}
\label{fig:fig4}
\end{center}
\end{figure}
\paragraph{The 1-loop 2-point amplitude (1)}
We consider the contribution of the 1-loop 2-point amplitude with 
two external lines placed at the different side of the world-sheet.
In this case, we have only one diagram $G_2^{(1)}$ and the contribution of the diagram is given by
\begin{eqnarray}
{\cal F}^{G_{2}^{(1)}}(\Phi^1,\Phi^2)
&=&
 \sum_I \Big\langle 
\Delta \ket{V_I} \eta_I \star \Phi^1 ,\,  \Delta( \ket{V'_I} \tilde{\eta}_I 
\star \Phi^2) \Big\rangle
\nonumber\\
&& \hspace*{-2.5cm}=
\phi^1_{F^1}(p_1)\phi^2_{F^2}(p_2) 
\nonumber\\
&&\hspace*{-2cm} \times \left( 
\sum_I (-1)^{|\phi_{F}||\mbox{\scriptsize $\ket{V_I}$}|} 
\int\! \frac{d^{26}p}{(2\pi)^{26}}\, \Big\langle 
\Delta \ket{V_I,p}\star \ket{F^1, p_1}, \,
\Delta( \ket{V'_I, -p } \star \ket{F^2, p_2} ) \Big\rangle\right)
\end{eqnarray}
where $ |\phi_{F}| = |\phi^1_{F^1}| = |\phi^2_{F^2}|.$ 
Note that ${\cal F}^{G_{2}^{(1)}}(\Phi^1,\Phi^2)=0$ if 
$(-1)^{|\phi^1_{F^1}| + |\phi^2_{F^2}|}=-1$.

\paragraph{The 1-loop 2-point amplitude (2)}
We consider the 1-loop 2-point amplitude with two external lines placed at the same side. 
In this case, there are three relevant diagrams $G_2^{\rm (2,i)}$,
$G_2^{\rm (2,ii)}$ and $G_2^{\rm (2,iii)}$.
Each diagram is calculated as 
\begin{eqnarray}
{\cal F}^{G_{2}^{\rm (2,i)}}(\Phi^1,\Phi^2)
&=& \sum_I \Big\langle 
\Delta ( \Phi^1 \star \Phi^2) ,\,
\Delta \ket{V_I} \eta_I \star \ket{V'_I} \tilde{\eta}_I 
) \Big\rangle
\nonumber\\
&=&\phi^1_{F^1}\phi^2_{F^2}
\sum_I \Big\langle 
\Delta ( \ket{F^1} \star \ket{F^2}) ,\,
\Delta \ket{V_I} \star \ket{V'_I} 
) \Big\rangle
,
\\
{\cal F}^{G_{2}^{\rm (2,ii)}}(\Phi^1,\Phi^2)
&=& \sum_I \Big\langle 
\Delta ( \Phi^2 \star \Phi^1) ,\,
\Delta \ket{V_I} \eta_I \star \ket{V'_I} \tilde{\eta}_I 
) \Big\rangle
\nonumber\\
&=&
\phi^1_{F^1} \phi^2_{F^2} (-1)^{|\phi_{F}|}
\sum_I \Big\langle 
\Delta ( \ket{F^2} \star \ket{F^1}) ,\,
\Delta \ket{V_I} \star \ket{V'_I}  
) \Big\rangle,
\\
{\cal F}^{G_{2}^{\rm (2,iii)}}(\Phi^1,\Phi^2)
&=& \sum_I \Big\langle 
\Phi^1 \star \Delta \ket{V_I} \eta_I  ,\, \Delta (\ket{V'_I}  \tilde{\eta}_I 
 \star \Phi^2)
\Big\rangle
\nonumber\\
&=&
\phi^1_{F^1} \phi^2_{F^2} 
\sum_I 
(-1)^{|\phi_{F}|+\mbox{\scriptsize $\ket{V_I}$}}
\Big\langle 
\ket{F^1} \star \Delta \ket{V_I} ,\, \Delta (\ket{V'_I}  
 \star \ket{F^2})
\Big\rangle
.
\end{eqnarray}
The contribution of each diagram to the amplitude is given as the coefficient of
 $ \phi^1_{F^1} \phi^2_{F^2}$.

To summarize, the 1-loop 2-point amplitude $A^{\mbox{\scriptsize 1-loop}}_{\phi^1_{F^1} \phi^2_{F^2}}$ is given by summing up the contribution from the diagrams 
$G_{2}^{\rm (1)}$, $G_{2}^{\rm (2,i)}$, $G_{2}^{\rm (2,ii)}$ and $G_{2}^{\rm (2,iii)}$ as
\begin{equation}
\sum_{\{G_{2}^{\rm (1)}
G_{2}^{\rm (2,i)}, G_{2}^{\rm (2,ii)}, G_{2}^{\rm (2,iii)}\}} 
{\cal F}^{G_{2}}(\Phi^1,\Phi^2)
\;= \;\phi^1_{F^1} \phi^2_{F^2} \;A^{\mbox{\scriptsize 1-loop}}_{\phi^1_{F^1} \phi^2_{F^2}}.
\end{equation}

\section{Gauge invariance of the on-shell amplitudes}

By using the explicit representation of amplitudes given in the previous section, we discuss the properties of on-shell amplitudes and prove that any on-shell amplitude, {\it i.e.,} the amplitude with all external string fields 
satisfy $Q\Phi^a=0$, is determined independently of gauge choices of propagators.
We also show that if at least one of the external string field is exact state ($\Phi^a=Q\Lambda$) the amplitude vanishes.

\subsection{Tree amplitudes}

We consider the general $n$-point tree amplitudes.
For the discussion of gauge invariance of on-shell amplitudes, 
we in particular specify the set of diagrams whose external string fields are placed in a particular circular permutation. 
We write the contribution of such a set of diagrams to the amplitudes as 
\begin{equation}
{\cal A}^{\rm tree}_{\rm cyclic} ( \Phi^1, \Phi^2,\cdots, \Phi^n )_{\Delta}
= \sum_{\{ G^{\rm tree}_{n} \}} {\cal F}^{G^{\rm tree}_{n}}_{\rm tree}
( \Phi^1, \Phi^2,\cdots, \Phi^n)_{\Delta}
\end{equation}
where the sum in the right-hand side is taken for all the $n$-point diagrams with external fields placed in, e.g., counter-clockwise order of $\{\Phi^1, \Phi^2,\cdots, \Phi^n\}$.
Note that in this section, we explicitly specify the propagator $\Delta$ we use to calculate the amplitudes. 
From the definition, we have 
\begin{equation}
{\cal A}^{\rm tree}_{\rm cyclic} ( \Phi^1, \Phi^2,\cdots, \Phi^n )_{\Delta}
= {\cal A}^{\rm tree}_{\rm cyclic} ( \Phi^n, \Phi^1,\cdots, \Phi^{n-1} )_{\Delta}. 
\end{equation}
Note that the total $n$-point amplitude $A^{\rm tree}$ is given by summing up all permutation of $\{\Phi^a\}$ as
\begin{eqnarray}
{\cal A}^{\rm tree} (\Phi^{1}, \Phi^{2},\cdots, \Phi^{n}  )_{\Delta}
&=& \frac{1}{n} \sum_{\{ {a_1}, {a_2},\cdots, {a_n}  \}} 
{\cal A}^{\rm tree}_{\rm cyclic} 
( \Phi^{a_1}, \Phi^{a_2},\cdots, \Phi^{a_n}  )_{\Delta}.
\\
&=& \phi^{1}_{F^1} \phi^{2}_{F^2} \cdots \phi^{n}_{F^n} 
A^{\rm tree}_{\phi^{1}_{F^1} , \phi^{2}_{F^2} ,\cdots, \phi^{n}_{F^n}  }\,.
\end{eqnarray}

Now we show that we can represent 
${\cal A}^{\rm tree}_{\rm cyclic}(\Phi^{a_1}, \cdots, \Phi^{a_n}   )_\Delta$ explicitly as
\begin{equation}
{\cal A}^{\rm tree}_{\rm cyclic} ( \Phi^1, \Phi^2,\cdots, \Phi^{n} )_{\Delta}
= \Big\langle [[ \Phi^1, \Phi^2,\cdots, \Phi^{n-1}]]_{\Delta} ,\Phi^{n} \Big\rangle.
\label{eq:deftree}
\end{equation}
Here, $[[\Phi^1,\cdots \Phi^n]]_{\Delta}$ is an operation which gives a Grassmann even string field from the $n$ ordered set of Grassmann odd string fields for a given choice of propagators.
The explicit definition is given recursively as follows.
For $n=2$, the operation is defined independently of $\Delta$ as 
\begin{equation}
[[\Phi_1, \Phi_2]]_{\Delta} = \Phi_1 \star \Phi_2 .
\end{equation}
The definition for $n=k$ is given recursively from the definition for $2\le n \le k-1$ as 
\begin{eqnarray}
[[ \Phi^1, \Phi^2,\cdots, \Phi^k]]_\Delta
&=&
\Phi^1 \star \Delta [[ \Phi^2, \Phi^3,\cdots, \Phi^{k}]]_\Delta
+\Delta [[ \Phi^1, \Phi^2,\cdots, \Phi^{k-1}]]_\Delta \star \Phi^{k}
\nonumber\\
&& +\sum_{m=2}^{k-2} 
\Delta [[ \Phi^1, \Phi^2,\cdots, \Phi^{m}]]_\Delta \star
\Delta [[ \Phi^{m+1}, \cdots, \Phi^{k}]]_\Delta.
\label{eq:def[[]]}
\end{eqnarray}
For example, for $n=3$ and $n=4$, we have 
\begin{eqnarray} 
[[1,2,3]]_\Delta &=& 1\star \Delta(2\star 3) + \Delta(1\star 2)\star 3,
\\
{[[1,2,3,4]]_{\Delta}} &=& 1\star \Delta(\Delta(2\star 3)\star 4)
+ 1\star \Delta(2\star \Delta (3\star 4))
+ \Delta (1\star 2 ) \star \Delta(3\star 4)
\nonumber\\
&&
+ \Delta(\Delta(1\star 2)\star 3)\star 4 
+ \Delta(1\star \Delta(2\star 3))\star 4.
\end{eqnarray}
The operation $[[\cdots]]_{\Delta}$ gives all the ways of $n$ aligned objects to be completely parenthesized by $n-2$ pairs of parentheses. 
Thus each term in $[[\Phi^1,\cdots \Phi^n]]_{\Delta}$ corresponds to each of the relevant diagrams appearing in the $(n+1)$-point tree amplitude ${\cal A}^{\rm tree}_{\rm cyclic}$. 
This means that eq.(\ref{eq:deftree}) is in general satisfied. 
Note that there is a cyclic symmetry 
\begin{equation}
\Big\langle [[ \Phi^1, \Phi^2,\cdots, \Phi^{n-1}]]_{\Delta} ,\Phi^{n} \Big\rangle
=
\Big\langle [[ \Phi^2, \Phi^3,\cdots, \Phi^{n}]]_{\Delta} ,\Phi^{1} \Big\rangle.
\end{equation}
The number of terms of $[[ \Phi^1, \cdots, \Phi^n]]_{\Delta}$ (or the number of different $(n+1)$-point diagrams) is known to be given by the $(n\!-\!1)$-th Catalan number $C_{n-1}$ which is given by 
$$C_{n-1}= \frac{[2(n-1)]!}{(n-1)! \, n!}\;.$$ 

Now we give the two lemmas which are important for proving the gauge invariance of the tree amplitudes.
\begin{lemma}
For $n$ Grassmann odd string fields $\Phi^i$ with $Q \Phi^i = 0 $, 
\begin{equation}
Q [[ \Phi^1, \Phi^2,\cdots, \Phi^n]]_{\Delta} =0 .
\end{equation}
\label{lemma1}
\end{lemma}
{\bf (proof) :~} For $n=2$, $Q[[\Phi^1,\Phi^2]]=Q(\Phi^1\star \Phi^2)=0$ from the assumption
and the property of $Q$ for the star product: $Q(A\star B) = QA\star B + (-1)^{|A|}A\star QB$. 
If we assume the equation is satisfied for $2\le n \le k-1$, then from eq.(\ref{eq:def[[]]}) and the relation $\{\Delta, Q\}=1$ (abbreviated form of eq.(\ref{eq:propcomm})), 
\begin{eqnarray}
Q[[ \Phi^1, \Phi^2,\cdots, \Phi^k]]
&=&
-\Phi^1 \star  [[ \Phi^2, \cdots, \Phi^{k}]]
+ [[ \Phi^1, \cdots, \Phi^{k-1}]] \star \Phi^{k}
\nonumber\\
&&\hspace*{-4cm}  +\sum_{m=2}^{k-2}( 
[[ \Phi^1, \cdots, \Phi^{m}]] \star
\Delta [[ \Phi^{m+1}, \cdots, \Phi^{k}]]
-\Delta[[ \Phi^1, \cdots, \Phi^{m}]] \star
[[ \Phi^{m+1}, \cdots, \Phi^{k}]]).
\end{eqnarray}
By using eq.(\ref{eq:def[[]]}) again, we can show that the equation vanishes.
$\qed$

\begin{lemma}
For $n$ Grassmann odd string fields $\Phi^i$ with $Q \Phi^i = 0$, 
\begin{equation}
 [[ \Phi^1, \Phi^2,\cdots, \Phi^n]]_{\Delta} 
= Q \epsilon ([[ \Phi^1, \cdots, \tilde{\epsilon}\Lambda^a ,\cdots, \Phi^n]]_{\Delta})
\end{equation}
if $\Phi^a = Q \Lambda^a$ for ${}^\exists \Phi^a$. 
Here $\epsilon$ and $\tilde{\epsilon}$ are Grassmann odd parameters with $\epsilon\tilde{\epsilon}=1$.
\label{lemma2}
\end{lemma}
{\bf (proof) :~} We assume $a=1$, {\it i.e.,} $\Phi^1=Q\Lambda^1$. Then for $n=2$, 
$[[\Phi^1, \Phi^2]] = Q \Lambda^1\star \Phi^2  = Q \epsilon ([[\tilde{\epsilon}\Lambda, \Phi^2]])$.
If we assume the statement is satisfied for $2\le n \le k-1$, then, as in the case of lemma~1, we can prove the statement for $n=k$ by using the relations eq.(\ref{eq:def[[]]}) and $\{\Delta, Q\}=1$ appropriately. 
For the case of $\Phi^a = Q \Lambda^a$ with $a\ne 1$, we can prove the statement similarly.
$\qed$

From the lemma~2, we immediately show that any on-shell tree amplitude vanishes if at least one of the external states is unphysical (exact) state:
\begin{theorem}
For $n$ external string fields $\Phi^i$ with $Q \Phi^i = 0$, 
\begin{equation}
{\cal A}^{\rm tree}_{\rm cyclic} ( \Phi^1, \Phi^2,\cdots, \Phi^{n} )_{\Delta}=0
\end{equation}
if $\Phi^a= Q\Lambda$ for ${}^\exists \Phi^a$.
\label{theorem1} 
\end{theorem}
Furthermore, we have the following theorem ensuring the gauge invariance of the on-shell tree amplitudes.
\begin{theorem}
For $n$ string fields $\Phi^i$ with $Q \Phi^i = 0$,  
\begin{equation}
{\cal A}^{\rm tree}_{\rm cyclic} ( \Phi^1, \Phi^2,\cdots, \Phi^{n} )_{\Delta}=
{\cal A}^{\rm tree}_{\rm cyclic} ( \Phi^1, \Phi^2,\cdots, \Phi^{n} )_{\Delta_0}\,.
\end{equation}
\label{theorem2}
\end{theorem}
Here $\Delta_0=b_0/L_0 $ is the propagator for Siegel gauge and 
$\Delta$ is that for another arbitrary gauge.
The propagators $\Delta_0$ and $\Delta$ are related to each other by $\Delta=\Delta_0 +QX-XQ$ (abbreviated form of eq.(\ref{eq:deffD0D})).
The proof of this theorem is almost straightforward with the above two lemmas.
First, from the lemma~1, we have
\begin{equation}
[[\Phi^1, \Phi^2,\cdots, \Phi^{n-1} ]]_{\Delta_0+QX-XQ}
= [[\Phi^1, \Phi^2,\cdots, \Phi^{n-1} ]]_{\Delta_0+QX}
.
\end{equation}
Then, we expand the right-hand side with respect to the order of $QX$.
The terms including at least one $QX$ can be collected as the sum of the form $[[\Psi^1,\cdots,\Psi^k]]_{\Delta_0}$ with $k<n-1$ where 
$\Psi^i $ is equal to one of $\Phi^a$ or can be written as the form $QX[[ \Psi^{a_1},\cdots \Psi^{a_j}]]_{\Delta_0}$.
Since any such $[[\Psi^1,\cdots,\Psi^k]]_{\Delta_0}$ includes at least one $\Psi^i=QX[[\cdots]]$, it can be written as $[[\Psi^1,\cdots,\Psi^k]]_{\Delta_0}= Q [[\cdots]]$ 
from lemma~2.
Thus, 
\begin{equation}
[[\Phi^1, \Phi^2,\cdots, \Phi^{n-1} ]]_{\Delta}
= [[\Phi^1, \Phi^2,\cdots, \Phi^{n-1} ]]_{\Delta_0}
+ \mbox{$Q$-exact terms}
\end{equation}
and 
\begin{equation}
\Big\langle [[\Phi^1, \Phi^2,\cdots, \Phi^{n-1} ]]_{\Delta}, \Phi^n \Big\rangle
= \Big\langle [[\Phi^1, \Phi^2,\cdots, \Phi^{n-1} ]]_{\Delta_0}, \Phi^n 
\Big\rangle.  \quad \qed
\end{equation}

We thus have proved the gauge invariance of on-shell amplitudes (theorem~\ref{theorem1})
and the decoupling of the on-shell unphysical amplitudes for general tree amplitudes  (theorem~\ref{theorem2}).  
\subsection{Loop amplitudes}
Now we discuss the gauge invariance of general on-shell loop amplitudes 
and prove the same statements as the two theorems for tree amplitudes.

In the case of tree amplitudes, the essential properties for the proof of gauge invariance are the relations $\{\Delta, Q\}=1$ and $Q(A\star B) = QA\star B + (-1)^{|A|}A\star QB$ satisfied for any string fields $A$ and $B$. 
With these properties, each $Q$ appearing in a diagram 
(from $\Phi^a=Q\Lambda$ or from $\Delta=\Delta_0+QX-XQ$) 
can be transmitted beyond the propagators and the star products in each diagram.
By summing up all the contributions from relevant diagrams, we see that the contributions from the diagrams which include at least one $Q$ cancel among them.

The cancellation properties used for loop diagrams are essentially the same as for the case of tree amplitudes explained in the previous subsection indirectly by using the $[[\cdots]]$ operation.
Note, however, that for loop amplitudes, we need to use the following additional property for the cancellation of diagrams: 
\begin{equation}
Q\,\left( \sum_{I} \Delta \ket{V_I}  \star \ket{V'_I} \right)
=
\sum_{I} \ket{V_I}  \star \ket{V'_I} 
= 0.
\label{eq:0tadpole}
\end{equation}
This property ensures that the operation of $Q$ on the `tadpole' 
diagram vanishes.
A formal proof of eq.(\ref{eq:0tadpole}) is given in \ref{app1}.
\begin{figure}[thbp]
\begin{center}
\begin{picture}(120,60)(0,0)
\thicklines
\put(65,30){\circle{30}}
\put(50,30){\line(-1,0){12}}
\put(80,20){$\Delta$}
\put(27,26){$Q$}
\put(110,26){\large$ =\,0$}
\end{picture}
\caption{{\bf eq.(\ref{eq:0tadpole}): $Q$ operation on the tadpole}}
\label{fig:fig5}
\end{center}
\end{figure}
We also summarize the structure of the cancellation properties of diagrams in \ref{app1}. Note that the similar argument has been given in ref.\cite{Kiermaier:2007jg}.

Consequently, we have the following general theorem which can be applied to any tree or loop amplitude. 
(In the case of tree amplitudes, the theorem reduces to theorems~\ref{theorem1} and~\ref{theorem2}.) 
\begin{theorem}
For $n$ external string fields $\Phi^i$ with $Q \Phi^i = 0$, 
\begin{equation}
{\cal A}^{\mbox{\scriptsize\rm $k$-loop}}_{\{G_T\}} ( \Phi^1, \Phi^2,\cdots, \Phi^{n} )_{\Delta}=
{\cal A}^{\mbox{\scriptsize\rm $k$-loop}}_{\{G_T\}} ( \Phi^1, \Phi^2,\cdots, \Phi^{n} )_{\Delta_0}.
\end{equation}
Furthermore, if $\Phi^a= Q\Lambda$ for ${}^\exists \Phi^a$,
\begin{equation}
{\cal A}^{\mbox{\scriptsize\rm $k$-loop}}_{\{G_{T}\}} ( \Phi^1, \Phi^2,\cdots, \Phi^{n} )_{\Delta}=0.
\end{equation}
\label{theorem3}
\end{theorem}
Here ${\{G_{T}\}}$ represents a set of all diagrams which have the same world-sheet topology with $n$ external fields and 
${\cal A}^{\mbox{\scriptsize\rm $k$-loop}}_{\{G_T\}}$ represents the sum of the contributions from all the diagrams in $\{G_T\}$.
For the tree amplitudes ($k=0$), ${\{G_{T}\}}$ consists of diagrams with external fields in a certain cyclic order and the statement obviously coincides with theorem~1 and 2.
For the one-loop amplitudes, ${\{G_{T}\}}$ is specified by the numbers and cyclic orders of fields at the two boundaries of the world-sheet.
For example, for 1-loop 2-point diagrams, 
${\{G_{T^2}\}}=\{G_{2}^{\rm (2,i)}, G_2^{\rm (2,ii)}  ,G_2^{\rm (2,iii)}\}$ 
(or ${\{G_{T^1}\}}=\{G_{2}^{\rm (1)}\}$) if the two external fields are at the same side (or the opposite side) of the boundary. 
Note that the $k$-loop $n$-point amplitude is given by summing up all the relevant 
${\cal A}^{\mbox{\scriptsize\rm $k$-loop}}_{\{G_T\}}$
as $\sum_T {\cal A}^{\mbox{\scriptsize\rm $k$-loop}}_{\{G_T\}}$.

The above theorem ensures that the on-shell amplitudes are consistently calculated 
for any consistent linear gauge fixing condition represented by eq.(\ref{eq:gaugecondn}). 
Furthermore, the on-shell amplitude with at least one unphysical state as an external state vanishes. 
Thus, when we calculate the on-shell amplitudes by using a particular gauge whose propagator is given by $\Delta$, we do not have to specify the gauges of external states, though we naturally have to use the same propagator $\Delta$ for all the internal lines.

\section{Summary and Discussions}
We have analyzed the perturbation theory of the cubic string field theory for general gauge fixing conditions represented by a linear equation ${\rm bpz}{\cal O}\Phi=0$ for the string fields $\Phi$. 
We have given a general prescription for obtaining the propagators $\Delta$ from the gauge fixed action $S_{\rm GF}$ for each gauge and have found the general properties of the propagators.
Explicitly, we showed that the propagators in general should satisfy the properties
$$
{\rm bpz}(\Delta^{\langle 4-n \rangle}) = \Delta^{\langle n \rangle},
\qquad
\Delta^{\langle n-1 \rangle}\Delta^{\langle n \rangle}=0,
\qquad
Q \Delta^{\langle n \rangle} + \Delta^{\langle n+1 \rangle}Q=1
$$
on the space of ghost number $n$ states.
Here,  $Q \Delta^{\langle n \rangle}  $ and $\Delta^{\langle n+1 \rangle}Q$ are projection operators of the space.
We also showed that the general form of the propagators is given by 
$$
\Delta^{\langle n \rangle}  =  
{\cal O}^{\langle n \rangle}  \, {\rm bpz}(\tilde{A}^{\langle -n+3 \rangle} ) 
Q \tilde{A}^{\langle n-1 \rangle}  \, {\rm bpz}({\cal O}^{\langle -n+4 \rangle} )
$$
where  $\tilde{A}^{\langle n \rangle}$ is in principle given as inverse of the operator  $T^{\langle n \rangle} (=Q {\cal O}^{\langle n \rangle} +{\rm bpz}{\cal O}^{\langle 3-n \rangle}Q$).
Then, by assigning the parameters with appropriate Grassmann  parity to 
the 2-string vertices $\ket{V_2}_{12}$, we fixed the possible sign ambiguities which might appear in the calculation of a Feynman diagram and established a consistent method for calculating the amplitudes. 
It can be said that the consistency of this method of calculation of amplitudes has been confirmed since we proved the gauge invariance of general on-shell amplitudes and the decoupling of on-shell unphysical amplitudes.

As we have explained in section~3,
the sign ambiguities we had to overcome for the calculation of amplitudes originated mainly from the fact that we have assigned the ghost number zero to the SL$(2,R)$ invariant vacuum $\ket{0}$. 
We may be able to give more straightforward method of calculation if we assign $-\frac{3}{2}$ to $\ket{0}$ from the beginning since this assignment of ghost number is natural from the viewpoint of the ghost number current. 
It would be instructive to establish an alternative method of calculating the amplitudes based on this assignment of ghost number and reinvestigate the properties of inner products or star products of string field theory in an algebraic point of view.

To prove the gauge invariance of on-shell amplitudes, we highly used the property $\{Q,\Delta \}=1$. We have to be careful that this relation can be applied for the state with $L_0\ne 0$. 
We may be lead to an incorrect result by an abuse of this relation. 
For example, if we carelessly inserted $1=\{Q,\Delta\}$ in front of $[[\Phi^1 , \cdots,\Phi^{n-1}]]_\Delta$ in eq.(\ref{eq:deftree}) in the case of $Q\Phi^i=0$ ($i=1,\cdots, n$), 
we would have the result
$$
 \Big\langle (Q\Delta+\Delta Q) [[ \Phi^1, \Phi^2,\cdots, \Phi^{n-1}]]_{\Delta} ,\Phi^{n} \Big\rangle
=
 \Big\langle Q (\Delta [[ \Phi^1, \Phi^2,\cdots, \Phi^{n-1}]]_{\Delta}) ,\Phi^{n} \Big\rangle
=0
$$
from lemma~1, and we would be lead to the wrong conclusion that any on-shell tree amplitude should vanish.

One reason why the propagators are not defined for $L_0\ne 0$
is that the gauge symmetry is not properly fixed in general for $L_0=0$.
For example, for Siegel gauge, this can be seen as follows. 
Suppose a state $\ket{f}_{n}$ satisfies the Siegel gauge condition $b_0 \ket{f}_{n}=0$ and 
the on-shell condition $L_0 \ket{f}_{n}=0$.
Then, the state $ Q \ket{f}_n$, if it is non-vanishing, satisfies the gauge condition $b_0 Q \ket{f}_n =0$ since $\{b_0,Q \}=L_0$.
This means that there exists a gauge transformation 
$\ket{f}_{n+1} \rightarrow \ket{f}_{n+1} + Q \ket{f}_{n}$
which transforms a state $\ket{f}_{n+1}$ within the bounds of the Siegel gauge condition.
Note that we considered the linear ($g=0$) part of the gauge transformation.

For general gauge conditions, the operator $T^{\langle n \rangle }$ defined by 
eq.(\ref{eq:Tn}) plays the same role as $L_0$ for Siegel gauge.
In fact, for a general gauge condition, if we have a state $\ket{f}_{n}$ satisfying 
${\cal O}^{\langle n \rangle} \ket{f}_{n}=0$ and 
the `on-shell condition' for the gauge $T^{\langle n \rangle }  \ket{f}_{n}=0$,
the gauge transformation $ \ket{f}_{n+1} \rightarrow \ket{f}_{n+1} + Q \ket{f}_{n}$ transforms a state $\ket{f}_{n+1}$ within the bounds of the gauge condition $ {\rm bpz}({\cal O}^{\langle 3-n \rangle}) \ket{f}_{n+1}=0$ since ${\rm bpz}( {\cal O}^{\langle 3-n \rangle}) Q \ket{f}_{n}=0 $.
Furthermore, the propagator $\Delta$ for general gauge given by eq.(\ref{eq:defprop1}) includes $\tilde{A}$ which is given as the `inverse' of the operator $T$ as in eq.(\ref{eq:TnAn}).

Thus we see that the operator $T^{\langle n \rangle }$ gives the on-shell condition corresponding to the gauge condition we take. 
The representation of $T^{\langle n \rangle }$ may differ according to the representation of ${\cal O}^{\langle n \rangle}$, e.g., 
$ {\cal O}^{\langle n \rangle} =  {\cal O}^{\langle n \rangle} c_0b_0$ or 
$ {\cal O}^{\langle n \rangle} =  {\cal O}^{\langle n \rangle} 
C^{\langle n-1 \rangle}  B^{\langle n \rangle}$ as explained in section~2.
The convenient choice of $C^{\langle n-1 \rangle} B^{\langle n \rangle}$ should be related to the choice of the convenient conformal frame for the gauge condition. 
To analyze the structure of $Q$-cohomology in a given conformal frame, 
it would be instructive to decompose $Q$ by using the appropriate basis of the state space in that frame as we write $Q$ using the modes $b_m$, $c_m$, and $L_m$ obtained by the expansion of $b(z)$, $c(z)$, and $T_{zz}(z)$ with respect to $z$. 

It would be interesting to have the world-sheet interpretation of propagators and of the amplitudes for general gauges.
Since the operator $T^{\langle n \rangle }$ plays the role of $L_0$ for each gauge, 
it may be natural to represent the propagator by using the Schwinger representation 
$\frac{1}{ T^{\langle n \rangle } } \sim \int_0^\infty ds e^{-s T^{\langle n \rangle }}.$
For such an analysis, it may be convenient to find a suitable representation of vertex operators for each gauge.

In this paper, we have proved the theorem which ensures the gauge invariance of on-shell amplitudes, which is the extremely natural property for a well-defined quantum field theory.
For a string field theory, however, we do not know the consistent method of defining the time coordinate and we have not succeeded in the canonical quantization.
Since we have shown that the general gauge invariance property holds for string field theory, we may expect that the other fundamental properties like the Ward-Takahashi identity are also found for string field theory.

\section*{Acknowledgements}
The work of M.K.~is supported in part by the Grants-in-Aid for Scientific Research (\#19540272) from the Japan Society for the Promotion of Science (JSPS).
\appendix 
\def\thesection{Appendix~\Alph{section}}
\renewcommand{\theequation}{\Alph{section}.\arabic{equation}}
\def\thesection{Appendix~\Alph{section}}
\setcounter{equation}{0}
\def\thesection{Appendix~\Alph{section}}
\section{Properties of the amplitudes}
\label{app1}
\def\thesection{\Alph{section}}
\subsection{Cancellation properties for the diagrams through the operation of $Q$}
We summarize the cancellation properties of different diagrams caused by the operation of $Q$. 
First, we note that the $Q$ operation on the star product and on the propagator is given by 
\begin{equation}
Q\epsilon (A\star B) = Q \epsilon A \star B +A\star Q\epsilon B
\label{eq:QeAB}
\end{equation}
and 
\begin{equation}
[Q\epsilon,\Delta]=   - \epsilon 
. 
\label{eq:Qedelta}
\end{equation}
Here we assign the Grassmann odd parameter $\epsilon$ after each $Q$ and treat $Q\epsilon$ as a single Grassmann even operator.
Then we can depict these operations without ambiguities of signs as in Figures~\ref{fig:figA1a} and \ref{fig:figA1b}. In the figures, the arrows mean the direction of the operation of $Q\epsilon$ and the dashed line means the collapsed propagator.  
\begin{figure}[htpb]
\begin{center}
\begin{picture}(300,60)(0,0)
\thicklines
\put(40,30){\line(-1,0){15}}
\put(40,30){\line(1,1){15}}
\put(40,30){\line(1,-1){15}}
\put(25,30){\circle*{4}}
\put(0,30){$Q\epsilon$}
\put(12,23){$\rightarrow$}
\put(57,5){$A$}
\put(57,45){$B$}
\put(90,26){$=$}
\put(150,30){\line(-1,0){15}}
\put(150,30){\line(1,1){15}}
\put(150,30){\line(1,-1){15}}
\put(135,30){\circle*{4}}
\put(167,5){$Q\epsilon A$}
\put(167,45){$B$}
\put(200,26){$+$}
\put(260,30){\line(-1,0){15}}
\put(260,30){\line(1,1){15}}
\put(260,30){\line(1,-1){15}}
\put(245,30){\circle*{4}}
\put(277,5){$A$}
\put(277,45){$Q\epsilon B$}
\end{picture}
\caption{{\bf $Q$ operation on the star product : eq.(\ref{eq:QeAB})}}
\label{fig:figA1a}
\end{center}
\end{figure}
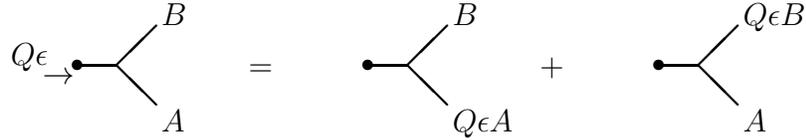
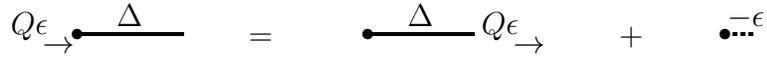
\begin{figure}[htpb]
\begin{center}
\begin{picture}(300,45)(0,5)
\thicklines
\put(25,30){\line(1,0){40}}
\put(25,30){\circle*{4}}
\put(40,33){$\Delta$}
\put(0,30){$Q\epsilon$}
\put(12,23){$\rightarrow$}
\put(90,26){$=$}
\put(135,30){\line(1,0){40}}
\put(135,30){\circle*{4}}
\put(150,33){$\Delta$}
\put(178,30){$Q\epsilon$}
\put(190,23){$\rightarrow$}
\put(230,26){$+$}
\put(270,30){\circle*{4}}
\put(270,30){\line(1,0){1.5}}
\put(273,30){\line(1,0){1.5}}
\put(276,30){\line(1,0){1.5}}
\put(279,30){\line(1,0){1.5}}
\put(271,33){$-\epsilon$}
\end{picture}
\caption{{\bf $Q$ operation on the propagator: eq.(\ref{eq:Qedelta})}}
\label{fig:figA1b}
\end{center}
\end{figure}

From these properties, we derive the relation which is satisfied for general Grassmann odd string fields $\Phi^A$, $\Phi^B$ and $\Phi^C$: 
\begin{eqnarray}
&& Q\epsilon \left( \Delta (\Phi^A \star \Phi^B) \star \Phi^C  
 +
\Phi^A \star \Delta (\Phi^B \star \Phi^C )\right)
\nonumber\\
&& \quad = \Delta (Q\epsilon \Phi^A \star \Phi^B) \star \Phi^C  
+ \Delta (\Phi^A \star Q\epsilon \Phi^B) \star \Phi^C  
+ \Delta (\Phi^A \star \Phi^B) \star Q\epsilon \Phi^C  
\nonumber\\
&& \qquad  + Q\epsilon\Phi^A \star \Delta (\Phi^B \star \Phi^C )
+ \Phi^A \star \Delta (Q\epsilon\Phi^B \star \Phi^C )
+ \Phi^A \star \Delta (\Phi^B \star Q\epsilon\Phi^C ).
\label{eq:Qest}
\end{eqnarray}
Note that we have used the cancellation of the two diagrams obtained after the collapse of propagator of each diagrams represented by the following trivial equation
\begin{equation}
\epsilon (\Phi^A \star \Phi^B) \star \Phi^C  
 +
\Phi^A \star \epsilon (\Phi^B \star \Phi^C )
=0.
\label{eq:est0}
\end{equation}
These properties are depicted in Figures~\ref{fig:figA2a} and \ref{fig:figA2b}.
\begin{figure}[htpb]
\begin{center}
\begin{picture}(420,240)(0,30)
\thicklines
\put(25,230){\line(1,0){60}}
\put(85,230){\line(1,1){15}}
\put(85,230){\line(1,-1){15}}
\put(50,230){\line(-1,1){15}}
\put(25,230){\circle*{4}}
\put(0,230){$Q\epsilon$}
\put(60,233){$\Delta$}
\put(12,223){$\rightarrow$}
\put(98,205){$\Phi^A$}
\put(98,247){$\Phi^B$}
\put(21,247){$\Phi^C$}
\put(125,226){$+$}
\put(175,230){\line(1,0){60}}
\put(235,230){\line(1,1){15}}
\put(235,230){\line(1,-1){15}}
\put(200,230){\line(-1,-1){15}}
\put(175,230){\circle*{4}}
\put(150,230){$Q\epsilon$}
\put(210,233){$\Delta$}
\put(162,223){$\rightarrow$}
\put(248,205){$\Phi^B$}
\put(248,247){$\Phi^C$}
\put(175,205){$\Phi^A$}
\put(10,146){$=$}
\put(35,150){\line(1,0){60}}
\put(95,150){\line(1,1){15}}
\put(95,150){\line(1,-1){15}}
\put(60,150){\line(-1,1){15}}
\put(35,150){\circle*{4}}
\put(70,153){$\Delta$}
\put(108,125){$Q\epsilon\Phi^A$}
\put(108,167){$\Phi^B$}
\put(31,167){$\Phi^C$}
\put(135,146){$+$}
\put(175,150){\line(1,0){60}}
\put(235,150){\line(1,1){15}}
\put(235,150){\line(1,-1){15}}
\put(200,150){\line(-1,1){15}}
\put(175,150){\circle*{4}}
\put(210,153){$\Delta$}
\put(248,125){$\Phi^A$}
\put(248,167){$Q\epsilon\Phi^B$}
\put(171,167){$\Phi^C$}
\put(275,146){$+$}
\put(315,150){\line(1,0){60}}
\put(375,150){\line(1,1){15}}
\put(375,150){\line(1,-1){15}}
\put(340,150){\line(-1,1){15}}
\put(315,150){\circle*{4}}
\put(350,153){$\Delta$}
\put(388,125){$\Phi^A$}
\put(388,167){$\Phi^B$}
\put(311,167){$Q\epsilon\Phi^C$}
\put(10,66){$+$}
\put(35,70){\line(1,0){60}}
\put(95,70){\line(1,1){15}}
\put(95,70){\line(1,-1){15}}
\put(60,70){\line(-1,-1){15}}
\put(35,70){\circle*{4}}
\put(70,73){$\Delta$}
\put(108,45){$\Phi^B$}
\put(108,87){$\Phi^C$}
\put(32,45){$Q\epsilon\Phi^A$}
\put(135,66){$+$}
\put(175,70){\line(1,0){60}}
\put(235,70){\line(1,1){15}}
\put(235,70){\line(1,-1){15}}
\put(200,70){\line(-1,-1){15}}
\put(175,70){\circle*{4}}
\put(210,73){$\Delta$}
\put(248,45){$Q\epsilon\Phi^B$}
\put(248,87){$\Phi^C$}
\put(175,45){$\Phi^A$}
\put(275,66){$+$}
\put(315,70){\line(1,0){60}}
\put(375,70){\line(1,1){15}}
\put(375,70){\line(1,-1){15}}
\put(340,70){\line(-1,-1){15}}
\put(315,70){\circle*{4}}
\put(350,73){$\Delta$}
\put(388,45){$\Phi^B$}
\put(388,87){$Q\epsilon\Phi^C$}
\put(315,45){$\Phi^A$}
\end{picture}
\caption{{\bf eq.(\ref{eq:Qest})}}
\label{fig:figA2a}
\end{center}
\end{figure}
\begin{figure}[htpb]
\begin{center}
\begin{picture}(300,80)(0,0)
\thicklines
\put(25,30){\line(1,0){25}}
\put(50,30){\line(1,0){1.5}}
\put(53,30){\line(1,0){1.5}}
\put(56,30){\line(1,0){1.5}}
\put(59,30){\line(1,0){1}}
\put(60,30){\line(1,1){15}}
\put(60,30){\line(1,-1){15}}

\put(50,30){\line(-1,1){15}}
\put(25,30){\circle*{4}}
\put(53,33){$\epsilon$}
\put(75,5){$\Phi^A$}
\put(75,47){$\Phi^B$}
\put(23,47){$\Phi^C$}
\put(110,26){$+$}
\put(155,30){\line(1,0){25}}
\put(180,30){\line(1,0){1.5}}
\put(183,30){\line(1,0){1.5}}
\put(186,30){\line(1,0){1.5}}
\put(189,30){\line(1,0){1}}

\put(190,30){\line(1,1){15}}
\put(190,30){\line(1,-1){15}}
\put(180,30){\line(-1,-1){15}}
\put(155,30){\circle*{4}}
\put(183,33){$\epsilon$}
\put(205,05){$\Phi^B$}
\put(205,47){$\Phi^C$}
\put(154,05){$\Phi^A$}
\put(240,26){$=\quad 0$}
\end{picture}
\caption{{\bf eq.(\ref{eq:est0})
}}
\label{fig:figA2b}
\end{center}
\end{figure}

We take the inner product between both sides of eq.(\ref{eq:Qest}) and another Grassmann odd string field $\Phi$ which is written as $\Phi=Q\Lambda = (Q\epsilon)(\tilde{\epsilon}\Lambda)$.
Here we insert two Grassmann odd parameters $\epsilon$ and $\tilde{\epsilon}$ 
with the property $\epsilon\tilde{\epsilon}=1$. 
Then the result becomes 
\begin{eqnarray}
&&\langle \Phi,  \Phi^A \star \Delta (\Phi^B \star \Phi^C) 
\rangle
+\langle \Phi,  \Delta (\Phi^A \star \Phi^B) \star \Phi^C 
\rangle
\nonumber\\
&& \; =
-\langle \tilde{\epsilon} \Lambda,  Q\epsilon\Psi^A \star \Delta (\Phi^B \star \Phi^C) \rangle
-\langle \tilde{\epsilon} \Lambda, \Psi^A \star \Delta (Q\epsilon\Phi^B \star \Phi^C) \rangle
-\langle \tilde{\epsilon} \Lambda, \Psi^A \star \Delta (\Phi^B \star Q\epsilon\Phi^C) \rangle
\nonumber\\
&& \quad
-\langle \tilde{\epsilon} \Lambda ,  Q\epsilon\Phi^A\star  \Delta(\Phi^B \star \Phi^C) \rangle
-\langle \tilde{\epsilon} \Lambda ,  \Phi^A\star \Delta( Q\epsilon\Phi^B \star \Phi^C) \rangle
-\langle \tilde{\epsilon} \Lambda ,  \Phi^A\star \Delta (\Phi^B \star Q\epsilon\Phi^C) \rangle.
\nonumber\\
&& 
\end{eqnarray}
The minus signs in front of all the terms in the right-hand side are due to the relation 
\begin{equation} 
{\rm bpz} Q = -Q \qquad ({\rm or}\qquad {\rm bpz} (Q\epsilon) = -(Q\epsilon)~  ). 
\end{equation}
Similarly, by taking the inner product between eq.(\ref{eq:est0}) and an arbitrary Grassmann odd string field $\Phi^D$, we have the relation 
\begin{equation}
\langle \Phi^A \star \Phi^B, \epsilon (\Phi^C \star \Phi^D) 
\rangle +
\langle \Phi^D \star \Phi^A, \epsilon (\Phi^B \star \Phi^C) 
\rangle
=0.
\label{eq:est0d}
\end{equation}
This is the essential relation which represents the structure of cancellation between s and t diagrams with propagators replaced by $-\epsilon$ caused by the operation of $Q$. (See Figure~\ref{fig:figA2c}.)
\begin{figure}[htpb]
\begin{center}
\begin{picture}(250,75)(0,-5)
\thicklines
\put(30,30){\line(1,0){1.5}}
\put(33,30){\line(1,0){1.5}}
\put(36,30){\line(1,0){1.5}}
\put(39,30){\line(1,0){1}}
\put(30,30){\line(-1,1){15}}
\put(30,30){\line(-1,-1){15}}
\put(40,30){\line(1,1){15}}
\put(40,30){\line(1,-1){15}}
\put(33,33){$\epsilon$}
\put(57,5){$\Phi^A$}
\put(57,44){$\Phi^B$}
\put(-1,5){$\Phi^D$}
\put(-1,44){$\Phi^C$}
\put(90,26){$+$}
\put(150,25){\line(0,1){1.5}}
\put(150,28){\line(0,1){1.5}}
\put(150,31){\line(0,1){1.5}}
\put(150,34){\line(0,1){1}}
\put(150,25){\line(-1,-1){15}}
\put(150,25){\line(1,-1){15}}
\put(150,35){\line(1,1){15}}
\put(150,35){\line(-1,1){15}}
\put(151,27){$\epsilon$}
\put(164,0){$\Phi^A$}
\put(164,49){$\Phi^B$}
\put(127,0){$\Phi^D$}
\put(127,49){$\Phi^C$}
\put(200,26){$=$~~  $0$}
\end{picture}
\caption{{\bf cancellation of two diagrams collapsed from s and t diagrams: eq.(\ref{eq:est0d})}}
\label{fig:figA2c}
\end{center}
\end{figure}

The above properties ensure that the contributions from the diagrams including 
the collapsed propagators by $Q$ are canceled in the set of diagrams which correspond to topologically equivalent world-sheet with insertions of external string fields. 
Note that the $Q$ in consideration is originated from the external field of the form $\Phi^a=Q\Lambda$ or from the difference between the propagator and that of the Siegel gauge $\Delta- \Delta_0 = QX-XQ$. 
Thus, for the on-shell amplitudes with all external fields vanishing with respect to $Q$ as $Q\Phi^i=0$, the contributions from the diagrams including $Q$ are completely cancelled among themselves and the statement of theorem~\ref{theorem3} is satisfied. 
Note that for loop amplitudes, eq.(\ref{eq:0tadpole}) is important for proving the theorem. 

\subsection{A proof of eq.(\ref{eq:0tadpole})}
We give a formal proof of the relation eq.(\ref{eq:0tadpole}):
$$
Q\,\left( \sum_{I} \Delta \ket{V_I}  \star \ket{V'_I} \right)
=
\sum_{I} \ket{V_I}  \star \ket{V'_I} 
= 0.
$$
Here, $\ket{V_I}$ and $\ket{V'_I}$ are the first and second components of the 2-string vertex $\ket{V_2}_{12}$ of eq.(\ref{eq:2verII'}).
The first equality is verified by using the properties eqs.(\ref{eq:QeAB}), (\ref{eq:Qedelta}) and (\ref{eq:prop2ver}).
The second equality is proved for example by applying the operators
$b_n-(-1)^n b_{-n}(\equiv B_n)$ and $c_n+(-1)^n c_{-n}(\equiv C_n)$ as ${\cal O}$ in the relation eq.(\ref{eq:prop2ver}). 
Since $B_n$ and $C_n$ act as derivative operators for the star product like $Q$ in eq.(\ref{eq:QeAB})~\cite{Rastelli:2000iu}, 
we have $B_n (\sum_{I} \ket{V_I}  \star \ket{V'_I}) =0$ and $C_n (\sum_{I} \ket{V_I}  \star \ket{V'_I})= 0$ for $n\in Z$. 
Thus $b_{-n}$ and $c_{-n}$ should appear in the combination of $(1+(-1)^nc_{-n}b_{-n})$ in $\sum_{I} \ket{V_I}  \star \ket{V'_I}$. Since $\sum_{I} \ket{V_I}  \star \ket{V'_I}$ has ghost number 3, we conclude $\sum_{I} \ket{V_I}  \star \ket{V'_I}= 0 $.



\end{document}